Vector electromagnetic theory of transition and diffraction radiation with application to the measurement of longitudinal bunch size

A.G. Shkvarunets and R.B. Fiorito

Institute for Electronics and Applied Physics
University of Maryland, College Park MD 20742

*Abstract*

We have developed a novel method based on vector electromagnetic theory and Schelkunoff's principles, to calculate the spectral and angular distributions of transition radiation (TR) and diffraction radiation (DR) produced by a charged particle interacting with an arbitrary metallic target. The vector method predicts the polarization and spectral-angular distributions of the radiation at an arbitrary distance from the source, i.e. in both the near and far fields, and in any direction of observation. The radiation fields of TR and DR calculated with the commonly used scalar Huygens model are shown to be limiting forms of those predicted by the vector theory and the regime of validity of the scalar theory is explicitly shown. Calculations of TR and DR done using the vector model are compared to results available in the literature for various limiting cases and for cases of more general interest. Our theory has important applications in the design of TR and DR diagnostics, particularly those that utilize coherent TR and DR to infer the longitudinal bunch size and shape. A new technique to determine the bunch length using the angular distribution of coherent TR or DR is proposed.

## Introduction

Optical transition radiation (OTR) from metallic targets is widely used for the measurement of transverse size, divergence and energy of electron and proton beams [1-5]. Recently the use of diffraction radiation for similar diagnostic purposes has been demonstrated [6-10]. Most accelerators and beam radiation devices produce *incoherent* TR and DR, e.g. at optical wavelengths which are usually much shorter than the bunch dimensions. Incoherent TR has the interesting property that when the radiating foil is large, i.e. the radiation parameter $\gamma\lambda/2\pi \ll a$ the size of the radiator, the angular distribution (AD) of the radiation is independent of the frequency of the emitted photon out to the plasma frequency of the radiating material. However, when $\gamma\lambda/2\pi \geq a$, TR can be considered, by application of Babinet's principle [6,11] to be a form of diffraction radiation and in this case the far field AD of the radiation is frequency dependent. Similarly, in the case of DR from an aperture, the AD is frequency dependent even when the radiating surfaces can be considered to be large [11]. As a result, for long wavelengths and/or at high energies, the far field angular distributions of DR and TR are both functions of the observed wavelength.

*Coherent* transition and diffraction radiation (CTR/CDR) are produced at wavelengths near and longer than the longitudinal bunch size. In the coherent regime, the spectral – angular density of the radiation has the well known form:

$$\frac{d^2 I_{Coh}}{d\omega d\Omega} = \frac{d^2 I_e}{d\omega d\Omega}\{N(N-1)S_T(\sigma_T, k, \theta)S_L(\sigma_L, \omega)\} \tag{1}$$

where the first term on the RHS of the equation is the single charge spectral-angular density, $N$ is the number of charges in the bunch and $S_T$ and $S_L$ are the transverse and longitudinal form factors of the bunch, respectively. In most cases the transverse form factor $S_T$, which depends on the transverse size and divergence of the beam, is close to unity and it is the longitudinal form factor $S_L$ that primarily determines the radiation production. This term is the squared Fourier transform of the longitudinal bunch distribution which, in principle, can be determined from the frequency spectrum of the radiation, provided some technique for retrieving the phase of the frequencies components can be developed.

The coherent spectrum be measured directly [12] or indirectly by means of autocorrelation interferometry [13,14]. RF LINACs commonly produce micro bunches with pulse durations (Δt) in the picosecond regime. In this case it is necessary to measure the spectrum in the FIR to mm wave band. For shorter bunches (Δt ~ 300 fs), e.g. those produced by laser-plasma interactions, the spectral content of the pulse extends to the THz regime. Both CTR and CDR in these wavelength bands have both been used to infer the longitudinal bunch size and attempts have been made to determine the temporal profile of the beam [14,15]. Because of the long wavelengths involved, the radiation factor $\gamma\lambda$ can easily exceed the size of the target used to generate the radiation even for beams with low to moderate energies. In this case, the finite size of the radiator introduces a frequency dependence into the spectral angular density of a single electron, i.e. the first term on the RHS of Eq.(1). This must be taken into account in order to correctly deduce the longitudinal bunch form factor from the measured spectrum.

The effect of the finite size of the radiator and the finite aperture of the detector or the transfer optics on the *spectrum* of CTR and CDR has been previously analyzed [15,16]. The main effects are a low frequency cutoff in the spectra of both TR and DR and, in the case of DR, an additional high frequency cutoff produced by the aperture, e.g. the slit width. Attempts to account for the low frequency loss have been made with limited success (cf. [17]).

Additionally, in all previous studies with the exception of one [16] the measurements are assumed to be made in the far field or wave zone, i.e. taken at $R >> \gamma^2\lambda$, a distance much larger than the coherence length of the radiation. At long wavelengths, the range of the actual measurements frequently violates this condition, i.e. measurements are commonly made in the near field or pre wave zone. In the near field the single electron spectral angular distribution has an additional frequency dependence, which depends on the distance to the source. This dependence must also be taken into account in the analysis of the spectral data. Also previous studies have considered only a few ideal source/radiator shapes, i.e. circular or rectangular apertures or foils (TR) and rectangular slits (DR) and the results are only applicable to high beam energies and/or normal incidence.

We call special attention to the case of off normal incidence. In all other analyses of TR and DR, to our knowledge, the effect of the inclination of the foil plays a minor role in the evaluation of the radiation field (see e.g. Ref. [6]). This is the direct result of the approximation used in the calculations, namely that the electron's fields, which are considered to be the source fields on the radiator, are taken to be purely *transverse* (i.e. radial) to the velocity of the particle. The scalar transverse component of the field of the electron is then integrated over the radiator surface (Huygens scalar field formulation) and used to calculate the radiation field at some distance from the source, either directly [18] or through intervening apertures or optics [19].

In this approximation the longitudinal component of the field, which is smaller than the transverse field by the Lorentz factor $\gamma$, is not taken into account. In addition, the vector nature of the radiation field is obscured, in the sense that the polarization of the radiation is, *a priori*, assumed to follow that of the electron, i.e. the source field. While these assumptions are approximately correct for high energy beams and/or normal incidence, they are generally invalid. They are particularly inapplicable for low energies and large inclination angles, large energies and small inclination angles and for small, ($r \leq \gamma\lambda$) and/or asymmetric foils. In these cases the transverse and longitudinal components of the field of the electron, are significant and must be taken into account in order to correctly predict the spectral angular distribution and polarization of the radiation.

In this paper we will develop a method based on electromagnetic theory which does not make any assumptions about the nature of the source field or the radiation field. The method can accurately calculate the spectral angular energy density of both DR and TR from an arbitrary metallic target or an aperture of arbitrary shape, in both the near and far fields, for any energy and inclination angle of the radiator. Such an approach provides the correct specular-angular distribution and polarization of the radiation fields which are not properly calculated in theories based on the scalar Huygens formulation. These results are essential to the design of diagnostics based on TR and DR, particularly those utilizing coherent TR and DR, to measure longitudinal bunch properties.

Our theoretical approach is based on Love's field equivalence theorem and one of Schelkunoff's field equivalence principles [20-22] which can be considered as the vector electromagnetic generalizations of Huygens's principle. The model uses only the spectral Fourier transform of the field of the electron, i.e. no spatial – wave number transforms are taken. We have used this method to calculate the angular distributions of TR and DR, i.e. for infinite and finite screens in various limiting cases where the distributions are theoretically well known, as well as in more general cases. We first show that the solutions for the far field and near field (pre wave zone) accurately match the available theoretical calculations using the scalar Huygens theory in the regime where this approach is applicable, i.e. high energy or normal incidence; second, we give an estimate of accuracy of our method in the near field zone; third, we theoretically compare the Huygens scalar and the vector solutions and demonstrate when the Huygens solution is valid; fourth, we apply the method to calculate the AD to situations where the Huygens scalar method is inapplicable; and finally, we will show that it may be possible use the AD of coherent TR and DR to infer the beam bunch length.

For completeness we mention a recent conference paper [23] which calculates TR using a vector diffraction approach developed by [24]. The latter applies Love's and Schelkunoff's principles to the diffraction of electromagnetic waves. We note, however, that [23] is a very preliminary analysis and is valid only in the limit of high electron energy. In comparison, our approach is complete, quite general and applies Schelkunoff's principles in a unique way to compute TR and DR from an arbitrary radiator.

**Theory**

*Coordinate systems*

In this work we use both Cartesian coordinates $(x, y, z)$, with unity vector triad $(\mathbf{i}, \mathbf{j}, \mathbf{k})$ traditional spherical coordinates $(r, \theta, \varphi)$ with unity vector triad $(\vec{e}_r, \vec{e}_\theta, \vec{e}_\varphi)$, where $x = r\sin\theta\cos\varphi$, $y = r\sin\theta\sin\varphi$, $z = r\cos\theta$ and cylindrical coordinates $(\rho, \varphi, z)$, $(\vec{e}_\rho, \vec{e}_\varphi, \vec{e}_z)$ with $x = \rho\cos\varphi$, $y = \rho\sin\varphi$. We designate the $(x, z)$ plane as the horizontal plane and the plane, which is parallel to and free to rotate around the $y$ axis, as the vertical plane (see Figure 1). We also utilize additional spherical coordinates, similar to the globe's latitude and longitude $(\theta_h, \theta_v)$ with unity vector diad $(\vec{e}_h, \vec{e}_v)$, where $\theta_h$ is the horizontal angle of observation in the vertical plane which measured from the axis $z$ to the axis $x$ (see Figure 1) and $\theta_v$ is the observation angle in the vertical plane measured from the horizontal plane $(x,z)$ to the $y$ axis. We describe the distribution of the intensity of radiation as a function of $\theta_h$ at given $\theta_v$ as a "horizontal scan", and the distribution of intensity as a function of $\theta_v$ at given $\theta_h$ as a "vertical scan". In most of the cases described below we assume that an electron with velocity $\vec{V}$ ($\vec{\beta} = \vec{V}/c$, where $c$ is the speed of light in vacuum) is incident on or emerges from the flat surface of an ideal conducting foil at the point $x = y = z = 0$. The orientation of the surface of the conductor is characterized by the unit normal vector $\vec{n}_S$.

*Previous Calculations of TR from an infinite conducting screen*

At normal incidence, $V_z = V$, $V_x = 0$, $V_y = 0$, and $n_{Sx} = 0$, $n_{Sy} = 0$, $n_{Sz} = 1$, the spectral angular intensity of transition radiation TR produced by the electron (forward and backward TR) is given by the familiar form [25]:

$$I = I_\theta = \frac{d^2W}{d\omega d\Omega} = \frac{e^2\beta^2}{\pi^2 c}\frac{\sin^2\theta}{(1-\beta^2\cos^2\theta)^2} \tag{2}$$

where $W$ is the radiated energy, $d\omega$ is the frequency band, $d\Omega = \sin\theta d\theta d\varphi$ is the solid angle, $e$ is the charge, and $\gamma$ is the Lorentz-factor. This radiation is symmetric about the z axis and the radiation field has a component in the $\vec{e}_\theta$ direction ($\vec{E} = E_\theta \vec{e}_\theta$ ) only. At $\sin^2\theta = \gamma^{-2}\beta^{-2} = (\gamma^2 - 1)^{-1}$ the intensity has maximum $I_{\max} = e^2\gamma^2 / 4\pi^2 c$. We use this value of the intensity as a normalization factor and refer to it as a unit of normal TR (NTR).

In other calculations of TR from an inclined conducting foil [26] the author has chosen a "tilted" trajectory angle for the electron, i.e. $V_x = V\sin\psi$, $V_y = 0$, $V_z = V\cos\psi$, $n_{Sx} = 0$, $n_{Sy} = 0$, $n_{Sz} = 1$, and the distributions of parallel and normal components of intensity are:

$$I_\parallel = \frac{e^2\beta^2}{\pi^2 c}\frac{\cos^2\psi(\sin\theta - \beta\cos\varphi\sin\psi)^2}{B^2} \tag{3}$$

$$I_\perp = \frac{e^2\beta^2}{\pi^2 c}\frac{\cos^2\psi(\beta\cos\theta\sin\varphi\sin\psi)^2}{B^2} \tag{4}$$

where $B = \left[1 - \beta(\sin\theta\cos\varphi\sin\psi - \cos\theta\cos\psi)\right]\cdot\left[1 - \beta(\sin\theta\cos\varphi\sin\psi + \cos\theta\cos\psi)\right]$.

Here the parallel and normal polarization are related to the plane of incidence, i.e. the (x,z) plane as above. Horizontal and vertical scans are calculated using the coordinate transformations: $\cos\theta = \cos\theta_h \cos\theta_v$, $\sin\varphi = \sin\theta_h / \sin\theta$, $\cos\varphi = \sin\theta_h \cos\theta_v / \sin\theta$.

Eqs. (3) and (4) can also be derived from the well known formulae of Pafomov [27] which are written in Cartesian coordinates ($x, y, z$), where the direction of the radiation is given by the unity vector $(\cos\theta_x, \cos\theta_y, \cos\theta_z)$. Pafomov's formulae can also be written in terms of the angles $\theta_v, \theta_h$ using the transformations: $\cos\theta_x = \cos\theta_v \sin\theta_h$, $\cos\theta_y = \sin\theta_h$, $\cos\theta_z = \cos\theta_v \cos\theta_h$.

*Method of Images Applied to Inclined Foils*

Transition radiation produced by an electron incident on or emerging from a tilted, flat, infinite perfect conductor can be calculated accurately and straight forwardly using the *method of image*s. Traditionally, TR is described in terms of radiation produced by the rapid stop or start of electron and its image charge [28]. In our paper we apply the image method to describe the production of TR from an inclined infinite foil but without requiring the stopping and rapid acceleration of the electron. In our picture the electron always moves forward with a constant velocity. However, the electron's positive image can abruptly change its direction of propagation, i.e. it "bounces" from the flat surfaces of the media (see Figure 2).

There are three stages of interaction of the electron with the conducting layer: (1) incidence of the electron on the first vacuum-metal interface, (2) motion of the electron inside the conductor and (3) emergence of the electron from the second interface. In the first stage the image charge moves in the direction of specular reflection with respect to the electron's direction. In the second stage the image moves with the electron in the same direction thus nullifying the field of the electron. In the third stage the image again moves in the direction of specular reflection with respect to the electron's direction. Since the velocity of the electron is constant it does not radiate. Rather, it is the image charge that radiates since its velocity changes discontinuously, first from the specular to the forward direction and then from the forward to the specular direction. In the Fraunhofer zone (far field), the radiation of a charged particle that sharply changes its direction of propagation is given by the well known Bremsstrahlung formula which, in the long wave approximation [29], is given by:

$$I = \frac{d^2W}{d\omega d\Omega} = \frac{e^2}{4\pi^2 c}\left|\frac{\vec{\beta}_2 \times \vec{n}}{1-\vec{\beta}_2 \cdot \vec{n}} - \frac{\vec{\beta}_1 \times \vec{n}}{1-\vec{\beta}_1 \cdot \vec{n}}\right|^2 \quad (5)$$

where $\vec{\beta}_1 = \vec{V}_1 / c$ , $\vec{\beta}_2 = \vec{V}_2 / c$ and $\vec{V}_1, \vec{V}_2$ are the velocities of the image charge before and after the bounce. For backward transition radiation (BTR) $\vec{\beta}_1 = \vec{\beta} - 2\vec{n}_S(\vec{\beta}\cdot\vec{n}_S)$ and $\vec{\beta}_2 = \vec{\beta}$. For forward transition radiation (FTR) $\vec{\beta}_1 = \vec{\beta}$ and $\vec{\beta}_2 = \vec{\beta} - 2\vec{n}_S(\vec{\beta}\cdot\vec{n}_S)$.

In the case of an electron moving parallel to the z axis ($V_x = 0$, $V_y = 0$, $V_z = V$) and a foil tilted in the horizontal plane ($n_{Sx} = \sin\psi$ , $n_{Sy} = 0$, $n_{Sz} = \cos\psi$), it follows from Eq.(1.5) that the horizontal and vertical components of the intensity of TR are given by

$$I_h = \frac{d^2W_h}{d\omega d\Omega} = \frac{e^2\beta^2}{4\pi^2 c}\left(\frac{\sin(\theta_h - 2\psi)}{1+\beta\cos\theta_v\cos(\theta_h - 2\psi)} + \frac{\sin\theta_h}{1-\beta\cos\theta_v\cos\theta_h}\right)^2 \quad (6)$$

$$I_v = \frac{d^2W_v}{d\omega d\Omega} = \frac{e^2\beta^2\sin^2\theta_v}{4\pi^2 c}\left(\frac{\cos(\theta_h - 2\psi)}{1+\beta\cos\theta_v\cos(\theta_h - 2\psi)} + \frac{\cos\theta_h}{1-\beta\cos\theta_v\cos\theta_h}\right)^2 \quad (7)$$

where $I_{h,v}$ are the horizontal (parallel to $\vec{e}_h$) and vertical (parallel to $\vec{e}_v$) polarization components. These equations are useful if a polarizer is used in the experiment. Formulae (5), and consequently (6), (7) are the benchmark solutions for TR from an infinite, flat metallic surface, i.e. they will be used to verify other models and approaches, including the method we present below.

*Vector Theory for TR and DR from an Arbitrary Conducting Screen*

Formulae ((3)-(7)) all describe both forward and backward transition radiation from a perfectly reflecting, infinite, flat, tilted foil observed in the Fraunhofer zone equally well. However, if the radiator has structure (i.e. holes, finite size, contours, etc.) with characteristic dimension $L \approx \beta\gamma\lambda / 2\pi \cos\psi$, then the radiation is different from TR from a flat, infinite foil and can be described as diffraction radiation (DR) [3].

We now present a general vector approach to the calculation of DR for an arbitrarily shaped screen inclined at an arbitrary angle with respect to the velocity vector of an electron with arbitrary energy. Our goal is to produce theory which can serve as a benchmark for the calculation of DR from a finite radiator, just as the image model provides for TR from an infinite radiator. As above, we first assume that the electron moves along axes $z$ ( $V_z = V$, $V_x = 0$, $V_y = 0$ ). In cylindrical coordinates $\rho, \varphi, z$ the Fourier components of the electric and magnetic fields of an electron moving in vacuum are

$$\vec{E}_e(\rho,\varphi,z,\omega) = \frac{e\alpha}{\pi V}\left(\vec{e}_\rho K_1(\alpha\rho) - \vec{e}_z \frac{i}{\gamma} K_0(\alpha\rho)\right)\exp(ik_e z) \quad (8)$$

$$\vec{B}_e(\rho,\varphi,z,\omega) = \vec{e}_\varphi \frac{\beta e\alpha}{\pi V} K_1(\alpha\rho)\exp(ik_e z) \quad (9)$$

where $K_0(\alpha\rho)$ and $K_1(\alpha\rho)$ are the zero and first order MacDonald functions, $\alpha = \omega / V\gamma$ and $k_e = \omega / V$.

In the Weissacker Williams approximation or the method of virtual photons [29,30], the Fourier components of the field of a relativistic electron are interpreted as plane electromagnetic waves each with a purely radial (transverse) electric field and a perpendicular, circular magnetic field of the same magnitude. This is a good approximation for high energy. However, in our model we assume that the electron has an arbitrary energy. Hence the Fourier components of the fields of the electron are not purely transverse and thus cannot be considered to be virtual photons in the traditional sense. We can, nevertheless, refer to the Fourier components of the field of an electron with arbitrary energy as waves with frequency $\omega$ and wave number $k_e$ propagating along the axis z with phase velocity $V$, but we note that these waves are *not* electromagnetic waves because electric field has a longitudinal component $E_{ez}$ which increases as the energy of the electron decreases. Moreover, the ratio of the magnetic to the electric field of these waves, $B_{e\varphi} / E_{er} \sim \beta$ decreases as the energy decreases.

We now consider how one of these waves interacts with a perfectly conducting boundary. The Cartesian components of the electric field (8) are:

$$E_{ex} = \frac{e\alpha}{\pi V} K_1(\alpha\rho)\cos\varphi \exp(ik_e z) \tag{10}$$

$$E_{ey} = \frac{e\alpha}{\pi V} K_1(\alpha\rho)\sin\varphi \exp(ik_e z) \tag{11}$$

$$E_{ez} = -\frac{i}{\gamma}\frac{e\alpha}{\pi V} K_0(\alpha\rho)\exp(ik_e z) \tag{12}$$

Assume that electron is incident on or emerges from the surface of a perfect conductor and induces the radiation of electromagnetic wave from the surface. The surface $S$ can have an arbitrary shape characterized by the unit vector function $\vec{n}_S(x_S, y_S, z_S)$ which is locally normal to the surface, at point $(x_S, y_S, z_S)$ and is directed into the vacuum. Since the fields inside conductor are zero the boundary conditions for the tangential component of the fields on the surface are given by

$$\vec{n}_S \times (\vec{E}_e + \vec{E}_S) = 0 \tag{13}$$

$$\vec{n}_S \times (\vec{B}_e + \vec{B}_S) = \frac{4\pi}{c}\vec{j}_e \tag{14}$$

where $\vec{E}_S$, $\vec{B}_S$ are the fields of the radiated electromagnetic wave and $\vec{j}_e$ is an induced surface electric current which is necessary to satisfy the boundary condition. From Eqs. (13) and (14) it follows that the tangential component of the electric field of the radiated wave is exactly defined by the electric field of the electron

$$\vec{n}_S \times \vec{E}_S = -\vec{n}_S \times \vec{E}_e \quad , \tag{15}$$

whereas the magnetic component

$$\vec{n}_S \times \vec{B}_S = \frac{4\pi}{c}\vec{j}_e - \vec{n}_S \times \vec{B}_e \tag{16}$$

has an uncertainty due to the uncertainty of the surface current $\vec{j}_e$. Note that the magnetic field of the radiated electromagnetic wave is dictated by the electric field of this wave rather than by the magnetic field of the electron, $\vec{B}_e$.

We assume that the radiated wave propagates from the surface into a vacuum. On the surface the distribution of the tangential component of the electric field of this wave is given by (15), but the distribution of the tangential component of the magnetic field is unknown because $\vec{j}_e$ is not known. Fortunately, the radiation from a surface with a known distribution of electric but unknown distribution of magnetic field can be calculated using Schelkunoff's field equivalence principles. Here we follow the formulation of Love's theorem and Schelkunoff's principles as given in [20].

According to Love's field equivalence theorem the further propagation of the primary electromagnetic wave which is incident on an imaginary surface is equivalent to the termination of this wave on this surface and the radiation of a secondary wave by a virtual surface magnetic current given by

$$\vec{j}_{Vm} = \frac{c}{4\pi}(\vec{n}_S \times \vec{E}_S) \tag{17}$$

and a virtual electric current given by

$$\vec{j}_{Ve} = \frac{c}{4\pi}(\vec{n}_S \times \vec{B}_S). \tag{18}$$

Both of these currents radiate downstream only. This formulation solves the so called ‘backward wave’ problem of Huygens’ scalar model which produces both forward and backward radiation from an arbitrary surface.

Schelkunoff modified Love’s theorem by introducing another virtual ideally reflecting surface adjacent to and upstream of Love’s virtual electric and magnetic current sheet. If the Schelkunoff surface is an ideal electric conductor then the image electric current induced on this surface is opposite to the virtual electric current $\vec{j}_{Ve}$ and the total radiating electric current is zero. At the same time the image magnetic current is equal to the virtual magnetic current $\vec{j}_{Vm}$ and the total radiating magnetic current doubles (see page 38 of [20]).

*In practice this means that it is enough to know the distribution of the electric field of the electromagnetic wave on the surface in order to calculate the further propagation of this wave.*

We now apply Schelkunoff's formulation directly to calculate transition and diffraction radiation from perfectly conducting surfaces. From Eq. 15, the source electric field on the surface is just the negative of the electric field of the electron whose components are given by Eqs. (10), (11), (12). The radiating electric and magnetic fields are then calculable in terms of the magnetic vector potential knowing the magnetic surface current $\vec{j}_{Vm}$ defined above in Eq. (17). The vector potential and the radiation fields at point $\vec{R}$ are then given by

$$\vec{A} = \frac{2}{c}\int_S \vec{j}_{Vm} \frac{\exp(ikR_S)}{R_S} dS \tag{19}$$

$$\vec{E} = -(\nabla \times \vec{A}), \qquad \vec{B} = \frac{i}{k}(\nabla(\nabla \cdot \vec{A}) + k^2 \vec{A}) \tag{20}$$

where $R_S = \left|\vec{R} - \vec{r}_S\right|$ is the distance from $dS$ to the point $\vec{R}$, $\vec{r}_S$ is radius vector of $dS$ and $\vec{E}_S$ is a complex vector which includes the phase $k_e z$ of the field of the electron.

In the far zone ($R_S \approx R \to \infty$, $1/kR \ll 1$) the expressions (19) can be rewritten as

$$\vec{A} = \frac{\exp(ikR)}{R}\vec{A}_\Omega\ , \quad \vec{A}_\Omega = \frac{2}{c}\int_S \vec{j}_{Vm} \exp(i\vec{k} \cdot \vec{r}_S) dS \tag{21}$$

where $\vec{A}_\Omega$ is a slowly changing amplitude. Hence (20) can be reduced to

$$\vec{E} = -i(\vec{k} \times \vec{A})\ , \qquad \vec{B} = k^{-1}(\vec{k} \times \vec{E}) \tag{22}$$

where $\vec{k} \parallel \vec{R}$ is the vector wave number. The electric field multiplied by $R$ is given by

$$(\vec{E}R) = -i(\vec{k} \times \vec{A})R = -i\exp(ikR)(\vec{k} \times \vec{A}_\Omega) \tag{23}$$

and the polarization component normal to the observation plane which is characterized by the normal unity vector $\vec{n}_r$ is

$$(\vec{E}_\perp R) = \vec{n}_r(\vec{n}_r \cdot (\vec{E}R)) \tag{24}$$

The total and the polarization components of the spectrum angular energy density of the radiation are given by

$$I = \frac{d^2W}{d\omega d\Omega} = c(\vec{E}R)\cdot(\vec{E}R)^* \ , \quad I_\perp = \frac{d^2W_\perp}{d\omega d\Omega} = c(\vec{E}_\perp R)\cdot(\vec{E}_\perp R)^* \ , \qquad I_{\|} = I - I_\perp \tag{25}$$

*Quasi Spherical Wave Approximation*

At a finite distance $R$ the radiation fields and the Poynting vector, which describes the energy flux, can be calculated using (19) and (20). Here we immediately confront the practical problem that the computation involving vector differential operators is very time consuming and possibly inaccurate and unstable. However, in many cases the distance to the observation point $R$ is moderate in the sense that the wave front is almost spherical, but the angular distribution of intensity is very different from the distribution at infinity (i.e. the far field distribution). In this case to simplify the calculations we can use expressions (19) and (22) to calculate the radiation fields. Formula (19) gives an exact solution for the magnetic vector potential and formula (22) gives an approximate, quasi spherical solution for the fields. The question is how much does the approximate solution (22) deviate from the exact one (20). To answer this question we assume that $R$ is large enough so that the deviation of the approximate solution from the exact solution is small. In this approximation using (19) , (22) and

$$\vec{E}_\perp = \vec{n}_r(\vec{n}_r \cdot \vec{E}) \,, \tag{26}$$

the spectrum of the energy flux and of its polarization components passing through the elementary area $ds$ of a sphere with radius $R$ are given by

$$J = \frac{dW}{d\omega ds} = c\vec{E}\cdot\vec{E}^* \ , \quad J_\perp = c\vec{E}_\perp\cdot\vec{E}_\perp^* \ , J_\parallel = J - J_\perp \tag{27}$$

Evidently the limit of solution (27) at $R \to \infty$ matches the exact solution (25) when the substitution $ds = R^2 d\Omega$ is made.

The deviation of the approximate solution (27) from the unknown exact solution can be found by estimating the terms neglected in (22) in comparison with (20). Using *local* Cartesian coordinates ($\hat{\mathbf{x}},\hat{\mathbf{y}},\hat{\mathbf{z}}$) with ($\hat{\mathbf{z}} \parallel \vec{R}$, $\vec{k} \parallel \hat{\mathbf{z}}$) and the origin at the observation point $\vec{R}$, the approximate components of the electric field (22) at the point $\vec{R}$ can be presented as

$$E_{\hat{x}} = ikA_{\hat{y}}\,, \qquad E_{\hat{y}} = -ikA_{\hat{x}}\,, \qquad E_{\hat{z}} = 0 \tag{28}$$

Here $A_{\hat{x},\hat{y},\hat{z}}$ are the exact components of the vector potential (19). These components can be presented in the form $A_{\hat{x},\hat{y},\hat{z}} = \tilde{A}_{\hat{x},\hat{y},\hat{z}}\exp(ikR)$, where $\tilde{A}_{\hat{x},\hat{y},\hat{z}}$ are slowly varying functions of coordinates. In this case the approximate intensity can be written as

$$J = c\left[\left|E_{\hat{x}}\right|^2 + \left|E_{\hat{y}}\right|^2\right] = k^2 c\left[\left|\tilde{A}_{\hat{x}}\right|^2 + \left|\tilde{A}_{\hat{y}}\right|^2\right]. \tag{29}$$

At the same time taking into account that $\partial/\partial\hat{z} = \partial/\partial R$ Eq. (20) yields the exact components of the field:

$$\mathcal{E}_{\hat{x}} = \frac{\partial A_{\hat{z}}}{\partial\hat{y}} - \frac{\partial A_{\hat{y}}}{\partial\hat{z}} = \left(-ik\tilde{A}_{\hat{y}} + \frac{\partial\tilde{A}_{\hat{z}}}{\partial\hat{y}} - \frac{\partial\tilde{A}_{\hat{y}}}{\partial R}\right)\exp(ikR)$$

$$\mathcal{E}_{\hat{y}} = \frac{\partial A_{\hat{x}}}{\partial\hat{z}} - \frac{\partial A_{\hat{z}}}{\partial\hat{x}} = \left(ik\tilde{A}_{\hat{x}} - \frac{\partial\tilde{A}_{\hat{z}}}{\partial\hat{x}} + \frac{\partial\tilde{A}_{\hat{x}}}{\partial R}\right)\exp(ikR)$$

$$\mathcal{E}_{\hat{z}} = \frac{\partial A_{\hat{y}}}{\partial\hat{x}} - \frac{\partial A_{\hat{x}}}{\partial\hat{y}} = \left(\frac{\partial\tilde{A}_{\hat{y}}}{\partial\hat{x}} - \frac{\partial\tilde{A}_{\hat{x}}}{\partial\hat{y}}\right)\exp(ikR). \tag{30}$$

The exact intensity can be written as $J_E = c\left[\left|\mathscr{E}_{\hat{x}}\right|^2 + \left|\mathscr{E}_{\hat{y}}\right|^2 + \left|\mathscr{E}_{\hat{z}}\right|^2\right] = J + \delta J$ . Using the complex number inequality $\left(\sum a_i\right)\cdot\left(\sum a_i\right)^* \le \sum (a_i \cdot a_i^*)$, the deviation of the approximate solution from the exact solution (30) can be estimated:

$$\delta J \le c\left[\left|\frac{\partial \tilde{A}_{\hat{x}}}{\partial R}\right|^2 + \left|\frac{\partial \tilde{A}_{\hat{y}}}{\partial R}\right|^2 + \left|\frac{\partial \tilde{A}_{\hat{y}}}{\partial \hat{x}}\right|^2 + \left|\frac{\partial \tilde{A}_{\hat{x}}}{\partial \hat{y}}\right|^2 + \left|\frac{\partial \tilde{A}_{\hat{z}}}{\partial \hat{x}}\right|^2 + \left|\frac{\partial \tilde{A}_{\hat{z}}}{\partial \hat{y}}\right|^2\right] \quad . \tag{31}$$

A further approximation of $\delta J$ can be made by replacing all components $\tilde{A}_{x,y,z}$ with $\tilde{A} = k^{-1}c^{-1/2}J^{1/2} \ge \tilde{A}_{x,y,z}$ (see Eq. (29)). If we define the relative deviation $D \equiv \delta J / J$ , then the maximum estimate of $D$ is

$$D_{\max} = \frac{1}{2k^2J^2}\left[\left(\frac{\partial J}{\partial R}\right)^2 + \left(\frac{\partial J}{\partial \hat{x}}\right)^2 + \left(\frac{\partial J}{\partial \hat{y}}\right)^2\right] \quad . \tag{32}$$

Note that the first term in Eq. (32) is the estimate of the terms $\sim k^{-2}R^{-2}$ which are neglected in far zone approximation. Indeed, if $J = J_0 R^{-2}$ then the $\partial / \partial R$ term is $2k^{-2}R^{-2}$. The derivatives are estimated numerically as

$$\partial J / \partial R \approx \left[J(\theta_h, \theta_v, R + \delta R) - J(\theta_h, \theta_v, R)\right] / \delta R \tag{33}$$

$$(\partial J / \partial \hat{x}) \approx R^{-1}\left[J(\theta_h + \delta\theta_h, \theta_v, R) - J(\theta_h, \theta_v, R)\right] / \delta\theta_h \tag{34}$$

$$(\partial J / \partial \hat{y}) \approx R^{-1}\left[J(\theta_h, \theta_v + \delta\theta_v, R) - J(\theta_h, \theta_v, R)\right] / \delta\theta_v \tag{35}$$

where $\theta_h, \theta_v$ are the angular coordinate of observation point and $\delta R$ , $\delta\theta_h$, $\delta\theta_v$ are the small but finite variations of the angular coordinates. The integral RMS deviation is given by

$$RMSD = \left( \left| \Delta\Omega \right|^{-1} \int_{\Delta\Omega} D_{\max}^{2} d\theta_h d\theta_v \right)^{1/2} \tag{36}$$

where $\Delta\Omega$ is the total solid angle of observation.

Depending on the wavelength observed, the position of a detector used to measure the radiated energy can be in the far field or in the near field zone. The estimate (36) is very useful because it helps to estimate the error in the calculation of the distribution of radiation for a given detector distance. In addition, a small value of the RMSD guarantees that the detector is placed in a radiation zone which has a well established spherical or quasi spherical wave front and flux of energy.

*Comparison of Vector and Scalar Huygens Models*

Huygens principle is usually used in diffraction problems to calculate the distortion of the intensity of a ray at small deflection angles that is introduced by an obstacle. It is assumed that the ray can be described as a monochromatic scalar wave characterized by amplitude $u_0$ and the wave vector $\vec{k}_0$. Assume that there is a surface $S$ which intersects the primary wave. According to the "simple" original formulation of the scalar Huygens principle [31] the surface cancels propagation of the primary wave and the forward propagation of the wave is described as a radiation of secondary waves from the surface. The amplitude $U$ of the secondary wave at the point $\vec{R}$ can be found as

$$U = -\frac{ik}{2\pi} \int_S u_0 \cos\Psi \frac{\exp(ikR_S)}{R_S} dS \tag{37}$$

where $u_0$ is a complex amplitude of the primary-source field on the surface, $\cos\Psi = k_0^{-1}(\vec{n}_S \cdot \vec{k}_0)$, and $\vec{n}_S$ is a unity vector normal to the surface. In the case of an electromagnetic wave it is assumed as a "zero" approximation that each field component of the primary wave produces the corresponding field component of the secondary wave.

If the primary wave is the Fourier component of the field of the electron moving along the z axis, then it is assumed that the secondary radiation is produced by the field adjacent to the solid part of the foil. The polarization components of the intensity are assumed to be the intensities of primary scalar waves produced by the corresponding polarized components $u_0 = u_x = -E_{ex}$ and $u_0 = u_y = -E_{ey}$ of the electric field of the electron. In the virtual photons approximation the longitudinal component of the field is entirely neglected, i.e. it is heuristically set equal to zero: $u_0 = u_z = -E_{ez} = 0$. Alternatively, if the magnetic field is used, $u_0 = u_x = \pm B_{ex}$ and $u_0 = u_y = \pm B_{ey}$ and the longitudinal component of the magnetic field of the electron is zeroed, i.e. $B_{ez}$=0. The virtual photon approximation is usually identified with the scalar Huygens model and we shall continue to refer to them in equivalent terms.

The scalar Huygens model is attractive from a computational point of view. Unfortunately there are limitations in applicability of this model. Firstly it is not clear how to take into account the longitudinal component of the electric field of the electron and secondly how the polarization of the source field is related to the polarization of the radiation field. These limitations can be better understood if one compares the fields derived by the scalar, Eq. (37) and vector, Eqs. (19), (22), models.

We can rewrite the vector formula for the radiated electric field as

$$\vec{E} = \vec{n}_S(\vec{k} \cdot \vec{F}) - \vec{F}(\vec{k} \cdot \vec{n}_S) \tag{38}$$

where $E_{ez}$ of electron is *included* and where

$$\vec{F} = -\frac{i}{2\pi}\int_S \vec{E}_S \frac{\exp(ikR_S)}{R_S} dS \tag{39}$$

is of the same form as Eq. (37). From (38) it follows that in the vector model each component of the radiation field is composed of all of the components of the source field. In contrast, i.e. in the scalar Huygens model, each component of the source field produces the same corresponding component of the radiating field. The question is there a case when the scalar model reproduces the accurate vector solution.

To answer this question, we compare the results of the vector and scalar models for the cases of normal incidence and inclined incidence on a flat infinite conductor in light of Eqs. (38) and (39). At normal incidence $\vec{n}_S = \vec{z}$ the components of the radiated field $E_{x,y,z}$ calculated using the vector model are given by

$$E_x = -F_x k_z \ , \ E_y = -F_y k_z \ , \ E_z = k_x F_x + k_y F_y \tag{40}$$

It is immediately obvious that the corresponding components of the electric field calculated from the scalar model ( cf. Eq. (37) ) do not match those of the vector model. Also it is clear from (39) and (40) that the component $E_{Sz}$ does not participate in radiation, thus making it reasonable to zero the longitudinal field as is done in scalar Huygens model. The intensity $J_V$ of the radiation computed from vector model can be written as:

$$J_V = J_S - c(F_x k_y - F_y k_x)^2 \tag{41}$$

where

$$J_S = ck^2 \left|F_x\right|^2 + ck^2 \left|F_y\right|^2 \tag{42}$$

is the intensity produced by the scalar approach. Thus the scalar intensity differs from the vector intensity by the term $\Delta J_{VS} = c(F_x k_y - F_y k_x)^2$. Note that this term equals zero in the case of an azimuthally symmetric field with an arbitrary radial distribution, a purely radial polarization and centroid on the $\vec{z}$ axis (the direction of incident particle). This is the case of TR or DR from azimuthally symmetric target such as a disc with a concentric circular hole and/or circular annuli. Thus if the calculation of the spectral- angular distribution of TR or DR is done based on the electric field of particle, then the intensity calculated from the scalar model exactly equals that of the vector model at all observation angles and all energies of the electron. In this case, however, there is a problem with the determining the polarization ( $\vec{E}$ is not normal to $\vec{k}$ ). On the other hand, if in the magnetic field is used to calculate the intensity using the scalar model, then there is no problem with the polarization ($\vec{E} \perp \vec{k}$), but the intensity has to be heuristically divided by $\beta^2$ in order to match the prediction of the vector model.

Now if the foil is inclined, i.e. $\vec{n}_S = \vec{n}_x + \vec{n}_z$, the situation is changed. At small angles of observation $k_x, k_y \ll k_z \approx k$ the components of the radiating field are given by:

$$E_x = k(n_x F_z - n_z F_x) \ , \quad E_y = -k n_z F_y \ , \quad E_z = 0 \tag{43}$$

where $n_x = \sin\Psi$, $n_z = \cos\Psi$. In this case the $E_{Sz}$ component always participates in radiation (see Eq. (39)). As we can see from Eq. (43), this component affects the radiation along with the $E_{Sx}$ component. Therefore the intensity obtained from the scalar model never matches that of the vector model unless $E_{Sz} \equiv 0$, i.e. the limit of high energy of the electron. In this limit the scalar model intensity matches the vector one, i.e.

$$J_V = J_S = ck^2 (F_x^2 + F_y^2) \cos^2 \Psi \ . \tag{44}$$

*However, the problem remains that there is no way to correctly calculate the polarization components of the intensity using the scalar Huygens approach, especially in the case of an arbitrary radiator and/or non relativistic energy of the electron.*

### *Computational Considerations*

Our computational implementation of the vector and scalar models used to calculate of TR or DR achieves high speed and accuracy by implementing an azimuthally symmetric mesh with a singularity in the center conforming to the singularity of the field of the electron. The Fourier component of the electric field of the electron is azimuthally symmetric function with singularity at $\rho \to 0$, $E(\rho) \sim K_1(\alpha\rho) \sim \rho^{-1}$; hence the density of energy flux grows to infinity $E^2 \sim \rho^{-2}$. In the code the integration of the field over the surface of interface is done in cylindrical coordinates $\rho, \varphi, z$ using an azimuthally symmetric mesh $d\varphi, d\rho$ with angular cell $d\varphi = 2\pi / N_\varphi$, where $N_\varphi$ is an integer . The radial mesh is adjusted in a way to keep the area of the cell $dS = \rho d\rho d\varphi \sim \rho^2 d\varphi$. This is done in order to equalize the energy flux $E^2 dS$ through the cells as much as possible and to minimize the number of "empty" cells with very small energy. Accordingly the radial mesh is generated as $\rho_N = \rho_1 (1 + d\varphi)^N$ where $\rho_1$ is the minimum radius. As the result, this mesh allows an accurate integration with a reasonable number of cells (e.g. if $N_\varphi = 150$ and $0.001 \le \alpha\rho \le 10$ then $N \approx 220$, and $N \cdot N_\varphi \sim 3.3 \cdot 10^4$ cells) which makes it very practical to calculate TR, DR for any situation. Note that the "homogeneous", non adjusted mesh of the same accuracy would require $\sim 10^8$ cells which make calculation absolutely impractical and barely stable.

**Results**

In this section we present calculations of TR and DR using the models and formulae described above for various cases in order to elucidate their similarities and differences. All the calculations referred to as the "vector model" are done by the same code "Vector" which incorporates Equations (10), (11), (12), (15), (17), (19), (22), (26) and (27). Different cases are calculated by specifying all the relevant parameters including the geometry of the radiator and the distance to the observation point. The calculations referred to as the Huygens or scalar model are calculated similarly by another code "Scalar" which is based on Equations (10), (11), and (37).

Figure 3a. shows a comparison of the solution (Eq. (3) and (4)) with the solution obtained with the image charge model (Eqs. (6) and (7)). The Figure shows a horizontal $\theta_h$ scan, taken at $\theta_v = 0$, of the total (unpolarized) TR intensity generated by a low energy electron $(\gamma = 5)$ incident on a perfect conducting foil oriented at 45 degrees with respect to the velocity of the electron. Since the Pafomov formula uses a tilted trajectory and the image charge model uses a tilted foil, we have transformed the image model curve by reflecting it about zero degrees and shifting it by the appropriate angle (45 degrees) in order to more easily compare the results. Figure 3b shows a polar plot of the image charge solution, the electron velocity direction and the foil for reference purposes. Figures 3a and 3b clearly show that forward and backward TR from flat foil is mirror symmetric about the plane of foil.

Figure 4 shows comparison of the corresponding vertical scans of TR calculated using the two methods presented above. Figures 3a. and 4 show excellent agreement between the two models.

Figures 5 and 6 compare horizontal and vertical scans for the image charge, vector and scalar models for the same parameters as Figures 3 and 4. The agreement between the vector and the image models is excellent. But the scalar Huygens solution is quite different in amplitude from the vector and image model solutions; the differences are more pronounced in the case of the horizontal scan in comparison to the vertical scan.

Figures 7 and 8 compare the horizontal and vertical scans (respectively) for the image, vector and scalar models in the somewhat extreme case of a high energy particle ( $\gamma = 500$ ) incident at near grazing incidence ($\psi = 89.5$ degrees). Again the image and vector horizontal scan solutions match perfectly but the scalar Huygens solution has a different distribution than the two other models. The vertical distributions for vector and the image model also agree perfectly while the Huygens solution is a little lower in amplitude than the other solutions. Note that the relative difference between the Huygens and the vector vertical scans are about the same for all angles. These Figures also demonstrate the accuracy and stability of the code used in this extreme case.

We now compare our vector theory calculations of TR done in the near field to those of Verzilov [32] for an infinite conducting screen at normal incidence. Verzilov uses the standard method of virtual photons which assumes that only the transverse component of the electric field need be used as source terms for calculation of the radiation fields. This is the usual high energy approximation used in most all calculations of TR. What is different about Verzilov's work is that he determines the radiation at an *arbitrary distance* from the source. Thus the radiation intensity is determined in the near field (pre wave zone) as well as the far field. Moreover, he identifies the relevant parameter, i.e. the vacuum coherence or "formation" length, as the distance where the near field solution differs from the usual Fraunhofer solution for TR. He further shows that the angular distribution of TR is a strong function of the ratio $R = 2\pi L/(\gamma^2 \lambda)$, the distance to the source in units of the vacuum coherence length. The experimental verification of the frequency dependence of the AD of TR in the pre wave zone, albeit in the incoherent regime (i.e. $\lambda \ll c\Delta t$) has been presented by Castellano, et. al. [33].

Figure 9. presents Verzilov's solutions for TR [32] and our vector calculations for various distances from the source measured in terms of the parameter *R*. Following [32] the amplitudes are presented in normalized units $e^2\gamma^2/\pi^2 c = 4\cdot NTR$ (see page 6 for the definition of NTR). As is clear from Figure 9 the Verzilov's and our vector model calculations agree perfectly. Note that Verzilov does not discuss the accuracy of his approximation. Our calculations of RMSD for all distributions presented in Figure 9 are: $2.31\cdot10^{-9}$, $3.15\cdot10^{-7}$, $7.4\cdot10^{-5}$ for the distance parameter R = 10, 1 and 0.1 respectively. Down to the smallest distance the wave has a well established spherical front.

We now compare the result of vector theory and scalar Huygens theory for the case of a finite screen in both the wave and pre wave zones. Both Shulga, et. al. [34] and Xiang, et. al [35] have used scalar Huygens approach to calculate TR for a finite disk. We will compare our vector theory results to those of Xiang, et. al. because their results are in a clear form which simplifies the comparison.

Xiang, et.al. have calculated the effect of finite target size on the AD of TR from a circular disk observed in the wave zone (Figure 2 ref. [35]) and DR from circular aperture in an infinite metallic screen observed in the pre wave zone (Figure 9 ref [35]). As we have previously shown [36], from Babinet's principle DR from the aperture is directly related to TR from the complementary screen, i.e. a finite circular disk and can be employed to calculate the later. The calculations of [35] are performed using the method of virtual photons and an expansion of the phase in terms of the distance between the target and the observation point. The first term in the expansion is retained for the wave zone and the first two terms are retained for the pre wave zone. Since both of the calculations are done for normal incidence of an electron passing through the center of the disk or aperture, the radiator is azimuthally symmetric with respect to the field of the electron. Therefore, the analysis provided above in the Section: *Comparison of Vector and Scalar Huygens Models* following Eq. (42) , indicates that theoretically there should be no differences between the vector and scalar Huygens calculations.

In Figure 10 and 11 we present a direct comparison of the ADs calculated from vector theory with those of Xiang, et. al. for wave zone TR (Figure 2 [35] ) and pre wave zone DR (Figure 9 [35]). The comparisons show that the exact vector calculations closely agree with the Huygens calculations although there are some small quantitative differences which are more pronounced for the far field (Figure 10) in comparison to the near field (Figure 11). The discrepancies are most likely the result of the approximations used in [35] in the expansion of the phase term. The RMSDs for distributions presented in Figure 11 are: $4\cdot 10^{-7}$, $8.7\cdot 10^{-7}$, $2.04\cdot 10^{-6}$ for the distance parameters 7, 3 and 1, respectively.

Finally we compare our vector theory to the *heuristic method* devised by Naumenko [37] to calculate DR from a finite target. Naumenko points our that the radiation field can be deduced from an integral of the current density on the surface of any target accounting for phase differences at each point on the surface. He presents an unproven, heuristic representation of the current density and uses this to calculate the radiation. To test his formulation, he shows that his solution matches known solutions for near and far field TR [32] as well as far field DR [38] in the appropriate limits. He then applies his method to the calculation of DR in both the near and far fields of a finite disk and a rectangular radiator which is inclined with respect to the direction of the electron.

In Figure 12 we compare our vector theory with Naumenko's calculations of DR from finite sized disk ( radius $a = 0.5\ \gamma\lambda$) , inclined at 45 degrees, for a particle with Lorentz factor $\gamma = 1000$ observed at $\lambda = 0.001$mm, as a function of the distance from the source in units of $R = L/(\gamma^2\lambda)$. Following [36] the amplitudes are measured in normalized units $e^2\gamma^2/c = 4\pi^2\cdot NTR$ . Note that there are some small deviations between the two solutions for higher values of R. The RMSDs are: $1.25\cdot 10^{-7}$, $1.51\cdot 10^{-6}$, $4.12\cdot 10^{-6}$, $1.66\cdot 10^{-5}$ in descending order of R. Again, the spherical wave approximation is very good.

Figure 13 compares the calculations of Naumenko for a square target with dimension p measured in units of $a = p/\gamma\lambda$, with those calculated using the vector model. The calculations using both methods are very close for all cases.

**New Method for Measuring Bunch Length**

With the exact vector method in hand we can use it to provide or test the calculation of the spectral angular density of TR or DR for any situation. We now apply the vector method to the calculation of *coherent* radiation from an inclined finite foil with diameter *D = 50mm*, beam energy *E = 100 MeV* and an incidence angle $\Psi = 45$ degrees. The chosen observation distance *L = 0.5m* and the frequency range of the calculation is 20-2000 GHz. These are typical experimental parameters of interest for using CTR for the measurement of the bunch length of a picosecond micro-pulse.

Figure 14 shows the horizontal angular distributions at $\theta_v = 0$ of the radiation $J(\theta_h, \omega)$ calculated for nine different frequencies in the range of 25 to 1000 GHz. These are a few representative samples of the total "spectrum" of 123 distributions ($J(\theta_h, \omega)$) in the frequency band from 25 to 1000 GHz used in the calculations described below. For the 8 sample frequencies which lie in the band of 25 to 800 GHz shown in Figure 14 the RMSDs are respectively: $4.52 \cdot 10^{-2}$, $2.03 \cdot 10^{-2}$, $7.35 \cdot 10^{-3}$, $2.1 \cdot 10^{-3}$, $9.32 \cdot 10^{-4}$, $5.27 \cdot 10^{-4}$, $2.43 \cdot 10^{-4}$ and $1.34 \cdot 10^{-4}$. Note that the accuracy of spherical approximation increases with the frequency. One can see that within solid angle of observation, i.e. 0 to 0.2 mrad, the frequency dependence of AD can be described as a low frequency cut off due to the finite size of radiator. Also note that the ADs are complex functions of angle and frequency, which should be taken into account in any method used to measure the spectrum of radiation and consequently determine the bunch form factor.

Figure 15 presents the longitudinal bunch form factors, $S_L(\sigma_L, \omega)$ of single Gaussian longitudinal charge distributions with half amplitude widths: 1, 1.5 and 2 ps, respectively. Figure 15 shows the effective high frequency cutoffs on the frequency spectrum due to the finite bunch length. These spectrums are relevant to the AD calculations that follow.

Figure 16. shows the angular distributions corresponding to each bunch length, where each broad band AD is computed from the integral

$$\frac{dW(\sigma_L, \theta_h)}{d\Omega} = \int_{\omega_1}^{\omega_2} J(\theta_h, \omega) S_L(\sigma_L, \omega) d\omega \tag{45}$$

using single Gaussian form factors and frequency dependent angular distribution functions $J(\theta_h, \omega)$. Figure 16. shows that bunch lengths differing by 0.5 ps are easily distinguishable. These results indicate that it may be possible to use the broad band AD *alone* to determine the rms bunch width, eliminating the need and complexity involved in measurement of the spectrum of the radiation. We have taken a simple form for the pulse, i.e. a Gaussian, for illustration, but any assumed shape could be assumed. The point is that the resulting integration shown above in Eq. (45), produces an angular distribution which is sensitive to the bunch form factor and accordingly to the bunch length and longitudinal distribution.

In an actual experiment the detector response $D(\omega)$ and the transmission loss $T(\omega)$ due to intervening optical components e.g. the observation window and possibly air absorption, may affect the broad band angular distribution calculated using Eq. (45). These effects must be either mitigated by proper design of the experiment or measured and taken into account. In the latter case, the integrand of Eq. (45) can be modified to include them, i.e.

$$\frac{dW(\sigma_L,\theta_h)}{d\Omega} = \int_{\omega_1}^{\omega_2} J(\theta_h,\omega) S_L(\sigma_L,\omega) D(\omega) T(\omega) d\omega \quad . \tag{46}$$

Recent experimental data shows that the rms bunch lengths measured with this new technique assuming a single Gaussian form factor agree well with those obtained using independent measurements [39]. In our preliminary experiment the detector response was reasonably flat over the frequency range of interest and the transmission losses were small. Therefore Eq. (45) was adequate to fit the data (i.e. a 6 % overall rms deviation between measured and fitted AD curves) and provided rms bunch widths that were within 10% of independent measurements. The AD may also be sensitive to the detailed distribution of the pulse and thus is a possible diagnostic of the longitudinal bunch shape. However, further validating experimental data need to be taken and a comparative analysis of fits to the data for various model distributions must be done. These will be presented in a future publication.

## Conclusions

In order to correctly calculate the spectrum and angular distribution of transition or diffraction radiation for inclined, finite targets it is important to have a method which makes no assumptions about the energy and inclination of the target where the longitudinal component of the electron's field may play a role. We have developed a very general vector approach which is applicable to any conducting surface, i.e. finite, arbitrarily curved or shaped surface oriented at any inclination angle with respect to the velocity of the particle.

We have tested our vector method with that of the image charge model which is the most fundamental and accurate solution for far field TR from an infinite, flat, perfectly conducting surface, i.e. Eq. (5). We derived analytical Formulae (6) and (7)) using the image model, in order to conveniently verify other available solutions including our vector method.

We have compared the AD of TR and DR calculated from our vector model with other available models for targets with various shapes and inclination angles. We have found the vector method gives accurate solutions in all known situations where calculations are available and the models can be directly compared.

We have shown that the scalar Huygens model is inapplicable in particular TR and DR cases when the energy is low and/or the inclination angle is high i.e. the case of near grazing incidence. In such cases there are *noticeable differences* between the Huygens solution and correct solutions provided by our vector model, method of image and Pafomov's formula.

Furthermore, we have applied our vector method to calculate the angular energy distribution of coherent TR for finite radiators observed at a moderate distance from the source, i.e. $R < \gamma^2 \lambda$, a case of experimental interest for the determination of the bunch length using CTR and CDR. We have shown theoretically that the broad band AD of energy (intensity integrated over the frequency band relevant to the pulse duration) is sensitive to the rms bunch length and may be used to measure this quantity. Recent preliminary experimental data support this finding.

**Figure Captions**

Fig 1

Coordinate system used to calculate TR and DR from an inclined foil.

Fig 2

Schematic of conducting foil, directions of the incoming charge, the image charge and directions of TR generated, using the method of images.

Fig 3a

Horizontal scan $\theta_h$ at $\theta_v = 0$ ($\gamma = 5$, $\psi = 45^0$) of unpolarized intensity of TR. Solid line - Pafomov formula, dashed line - image model, dotted – transformed image line.

Fig 3b

Horizontal polar plot at $\theta_v = 0$ of image solution, showing the orientation of the foil and direction of the electron.

Fig 4

Vertical scan $\theta_v$ ( $\gamma = 5$, $\psi = 45^0$) of unpolarized intensity of TR; solid line: Pafomov formula at $\theta_h = \pi / 4$ (forward radiation) and at $\theta_h = 3\pi / 4$ (backward); dotted line: image charge model calculation at $\theta_h = 0$ (forward radiation) and at $\theta_h = -\pi / 2$ (backward).

Fig. 5

Comparison of horizontal scans of intensity of TR from an infinite screen inclined at angle $\Psi = 45$ degrees for an electron with Lorentz factor $\gamma = 5$, computed from image charge-solid, scalar Huygens-dash and vector-dots models.

Fig. 6

Comparison of vertical scans of intensity of TR from an infinite screen inclined at angle $\Psi = 45$ degrees for an electron with Lorentz factor $\gamma = 5$, computed from image charge-solid, scalar Huygens-dash and vector-dots models. In all cases $\theta_h = 0, -\pi / 2$.

Fig. 7

Comparison of horizontal scans of intensity of TR from an infinite screen inclined at angle $\Psi = 89.5$ degrees (near grazing incidence) for an electron with Lorentz factor $\gamma = 500$, computed from image charge-solid, scalar Huygens-dash and vector-dots models.

Fig. 8

Comparison of vertical scans of intensity of TR from an infinite screen inclined at angle $\Psi = 89.5$ degrees for an electron with Lorentz factor $\gamma = 500$, computed from image charge-solid, scalar Huygens-dash and vector-dots models. In all cases $\theta_h = 0, -1^0$.

Fig. 9

Comparison of calculation of TR from an infinite screen calculated in [32] and by vector theory for various distances from the source, measured in terms of the dimensionless parameter $R = 2\pi L/(\gamma^2\lambda)$, the ratio of the distance to the coherence length.

Fig. 10

Comparison of TR from a circular disk at normal incidence using the scalar Huygens formulation of [35] with vector theory for various values of the ratio of b, the radius of the disk, to $\gamma\lambda$.

Fig. 11.

Comparison of DR from a circular hole with radius $a = \gamma\lambda$ in an infinite metallic screen using the scalar Huygens formulation of [35] with vector theory for various distances $L$ to the source measured in terms of the ratio $R = L/\gamma^2\lambda$.

Fig. 12

Comparison of the calculation of DR from [36] and vector theory for an inclined (at 45 degrees) finite circular disk with radius $a = 0.5\gamma\lambda$, for an electron with Lorentz factor $\gamma$ = 1000 and observed wavelength $\lambda$ = 0.001mm, for various distances to the source measured in terms of $R = L/(\gamma^2\lambda)$.

Fig. 13

Comparison of the calculation of far field DR from [36] and the vector theory for an inclined (at 45 degrees) finite square plate with linear dimension $p \times p$ measured in units of $a = p / \gamma\lambda$, Lorentz factor $\gamma$ = 1000 and observed wavelength $\lambda$ = 0.001mm, for various values of $a$.

Fig. 14

Angular distributions of single electron DR for various frequencies in the range of 25 to 800 GHz for a 100 MeV beam, 50 mm disk, inclination angle of 45 degrees at distance of 0.5 meters from the source.

Fig. 15

Single Gaussian bunch spectra (longitudinal form factor) for various FWHMs: 1, 1.5 and 2 picoseconds.

Fig. 16

Angular distributions of Coherent TR from a 50 mm disk calculated from vector theory and a single Gaussian longitudinal beam distribution with full widths of 1, 1.5 and 2 picoseconds.

FIGURE 1

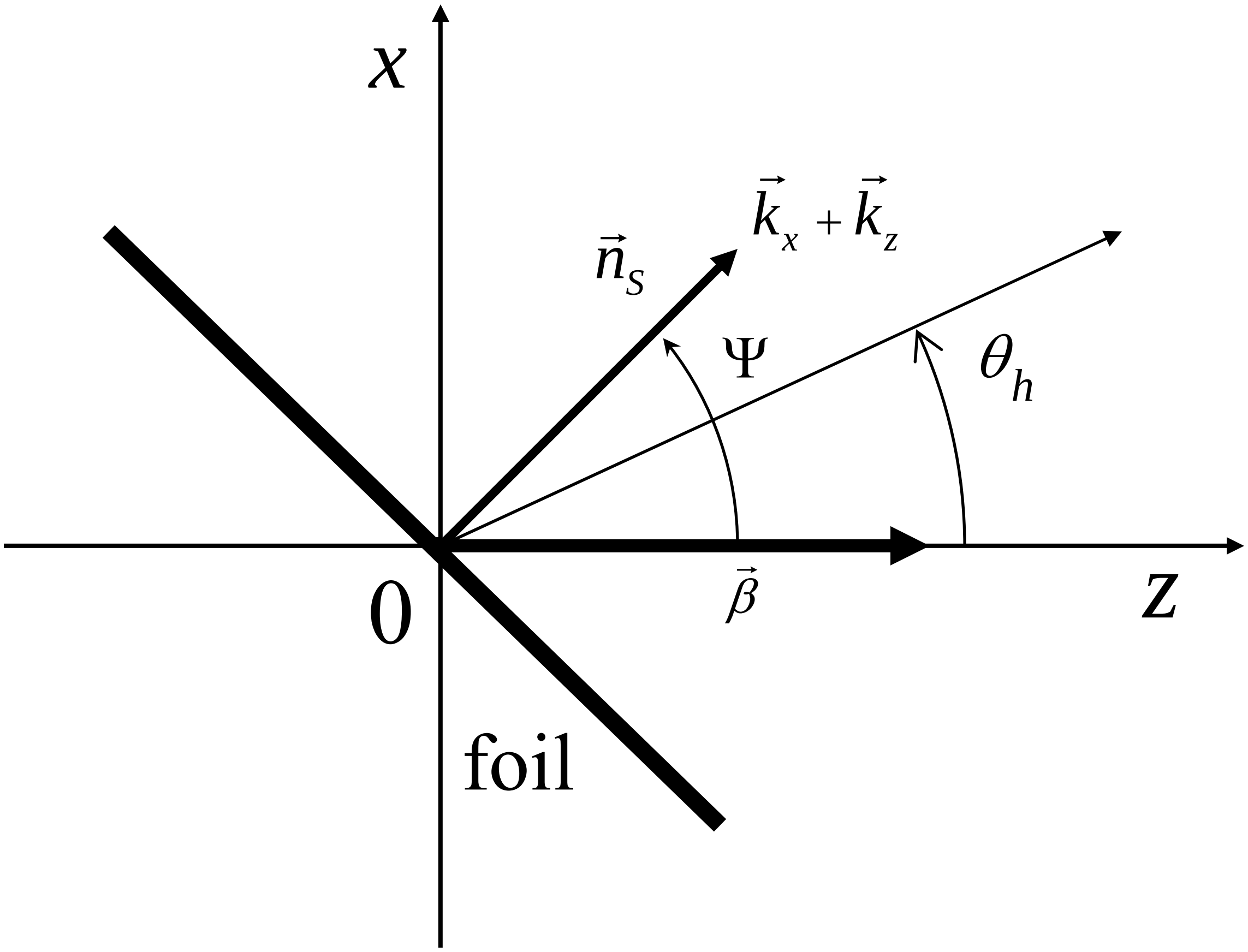

FIGURE 2

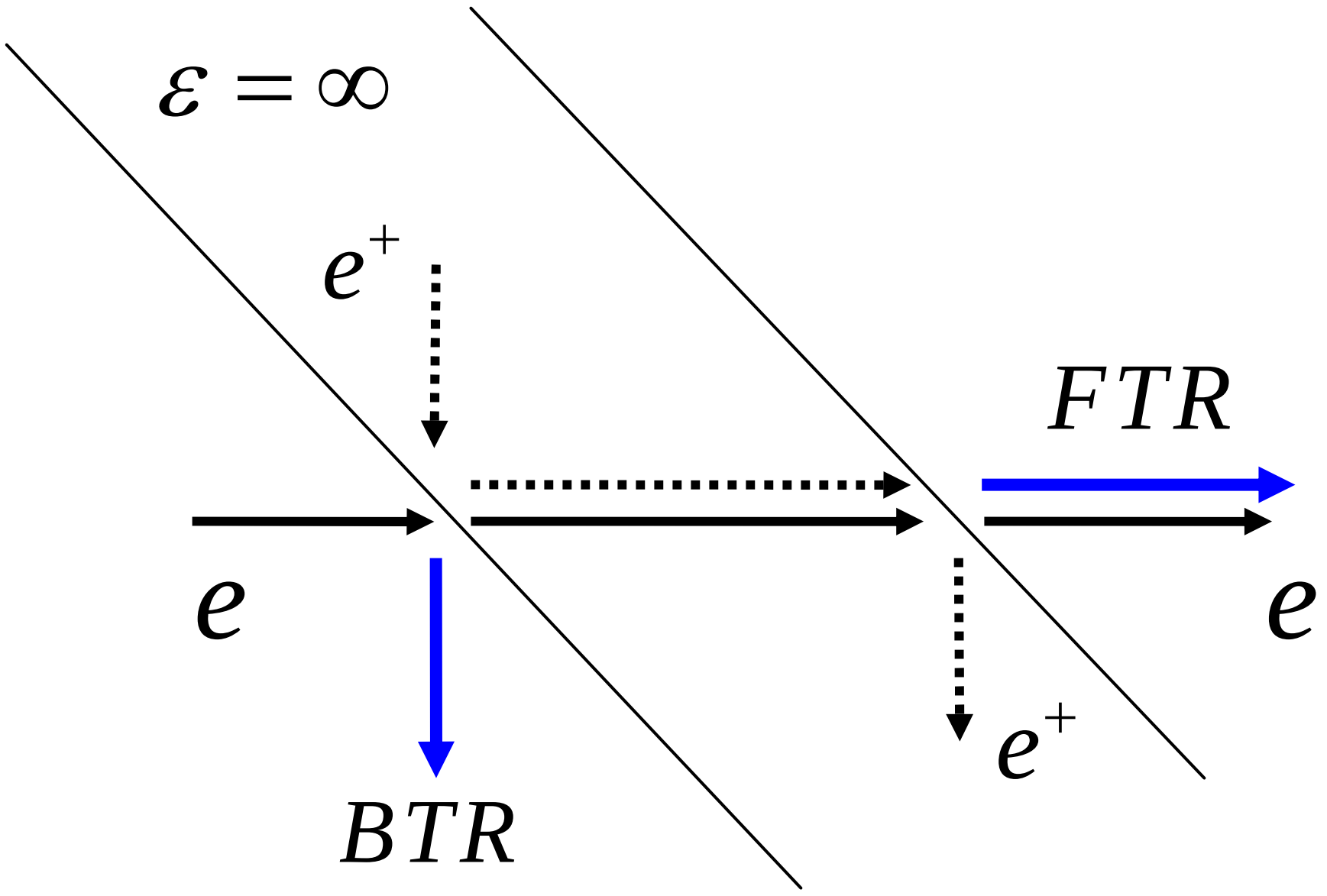

FIGURE 3

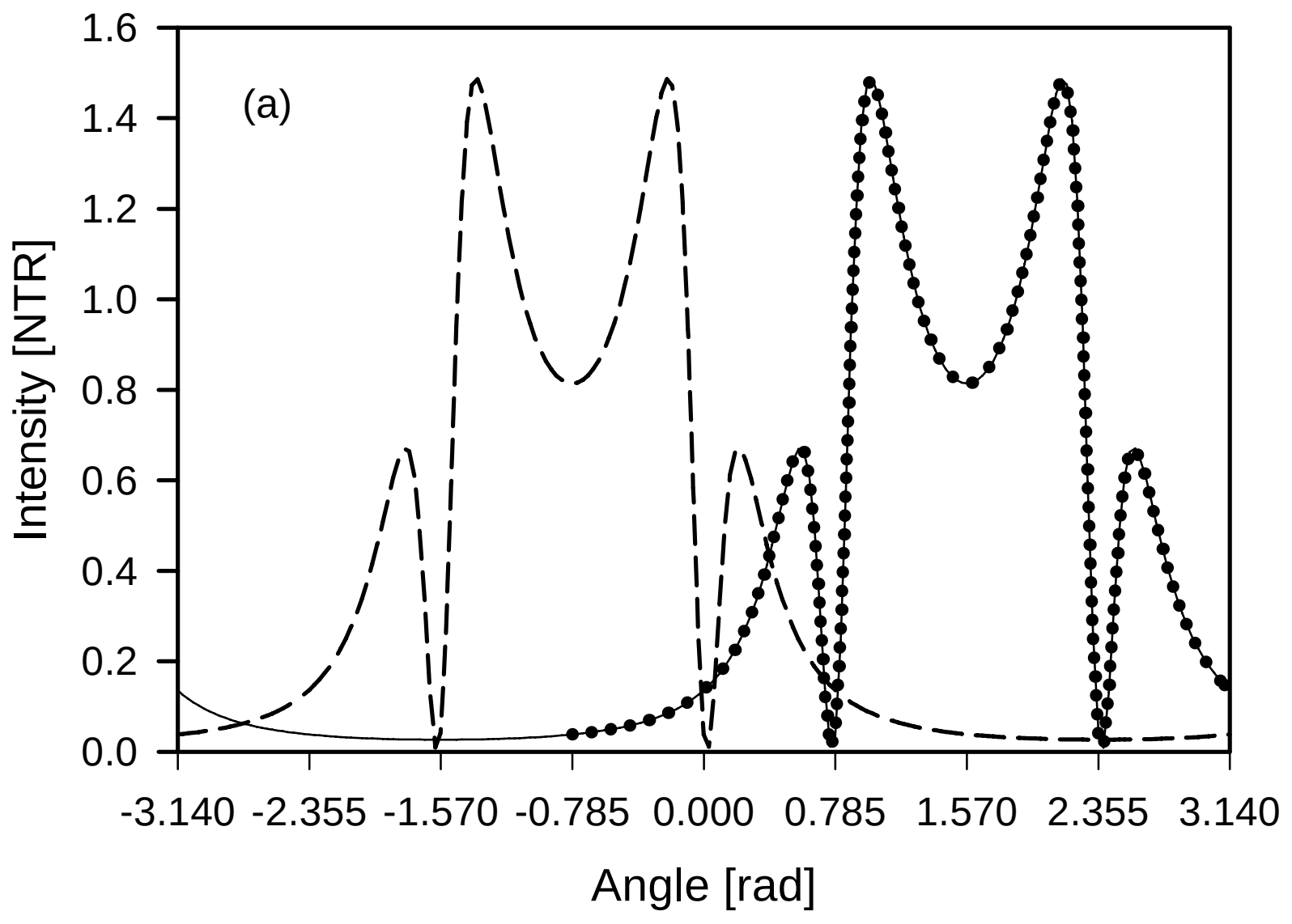

(a)
Intensity [NTR]
Angle [rad]
1.6
1.4
1.2
1.0
0.8
0.6
0.4
0.2
0.0
-3.140 -2.355 -1.570 -0.785 0.000 0.785 1.570 2.355 3.140

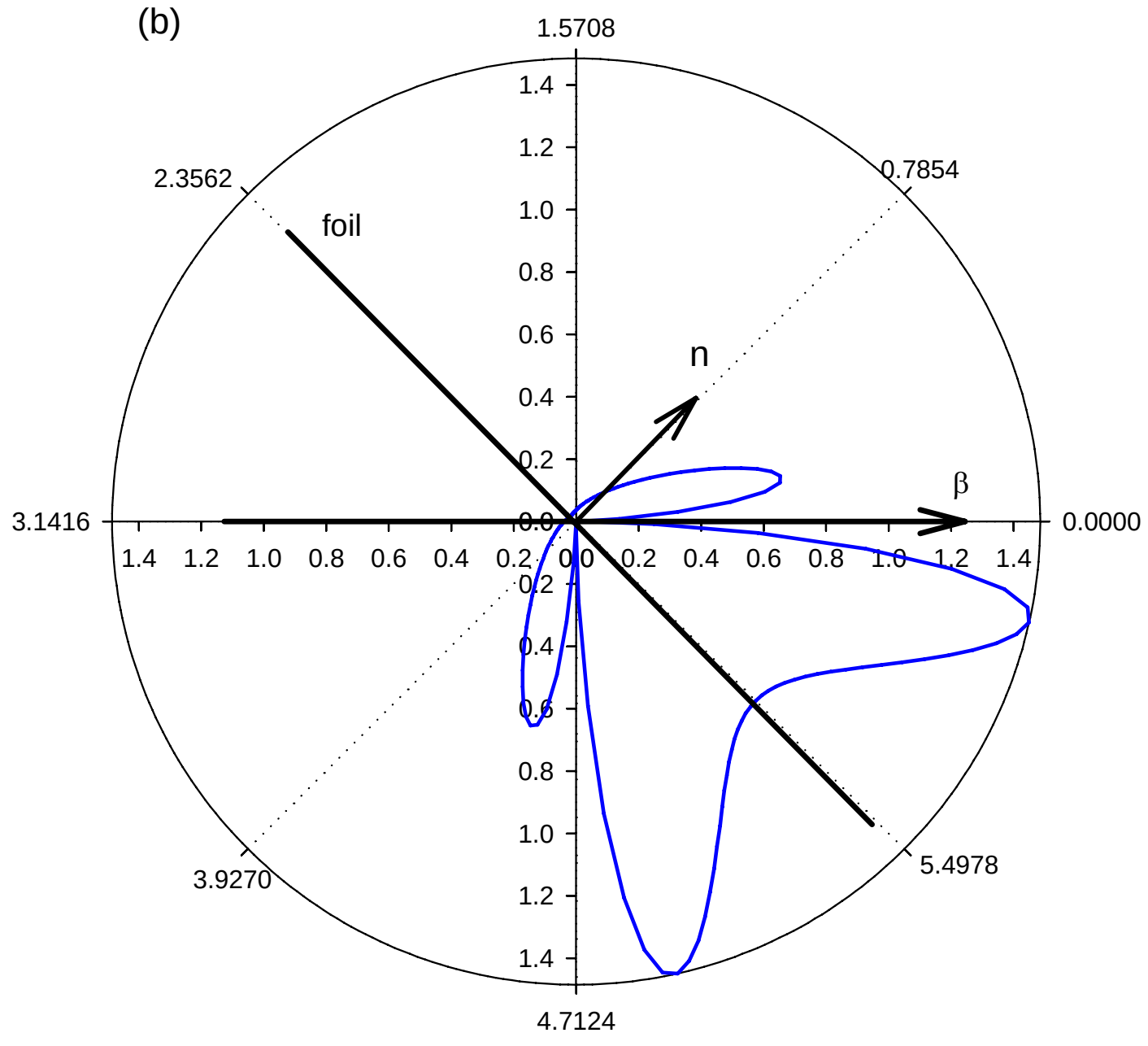

FIGURE 4

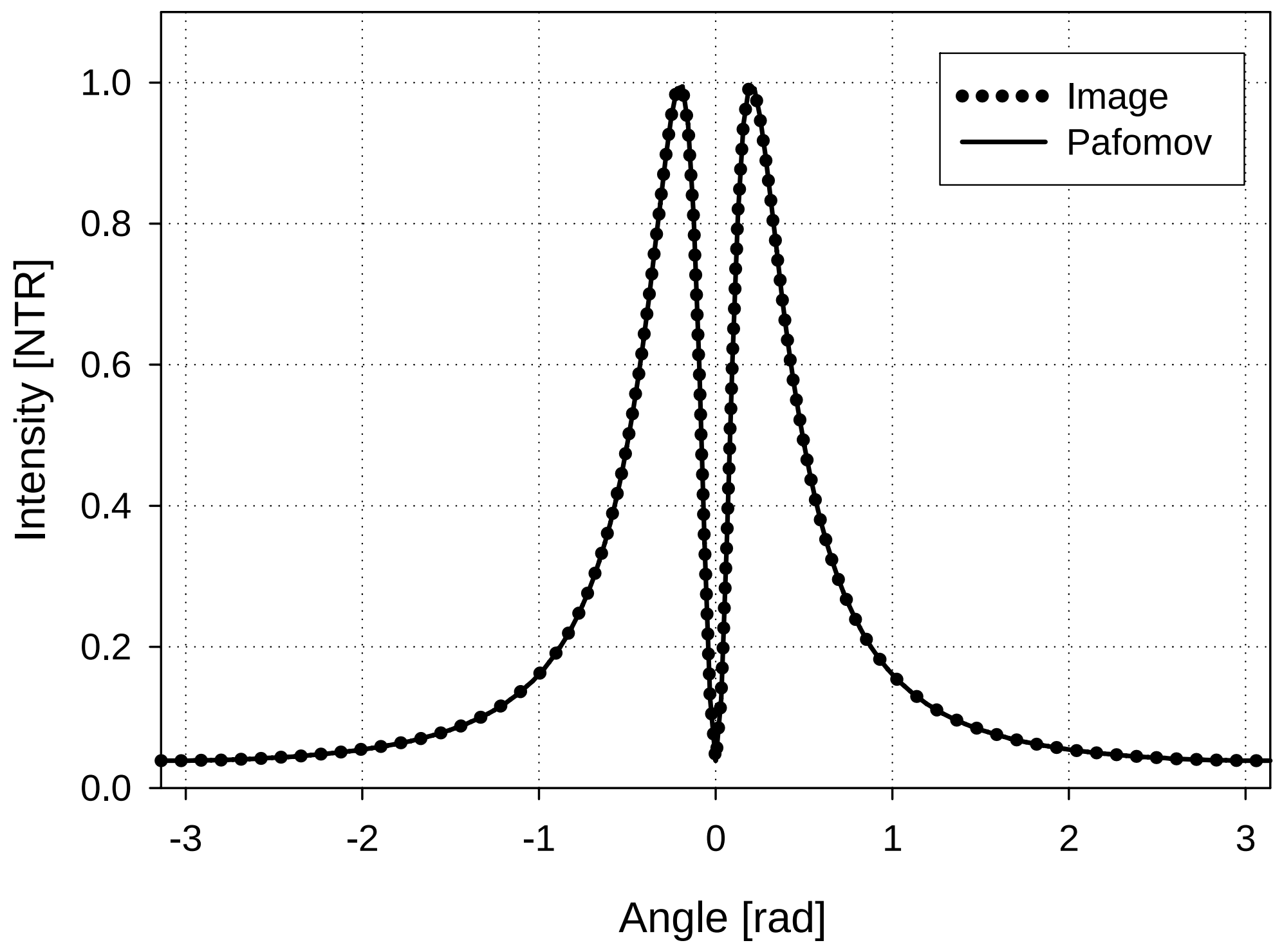

FIGURE 5

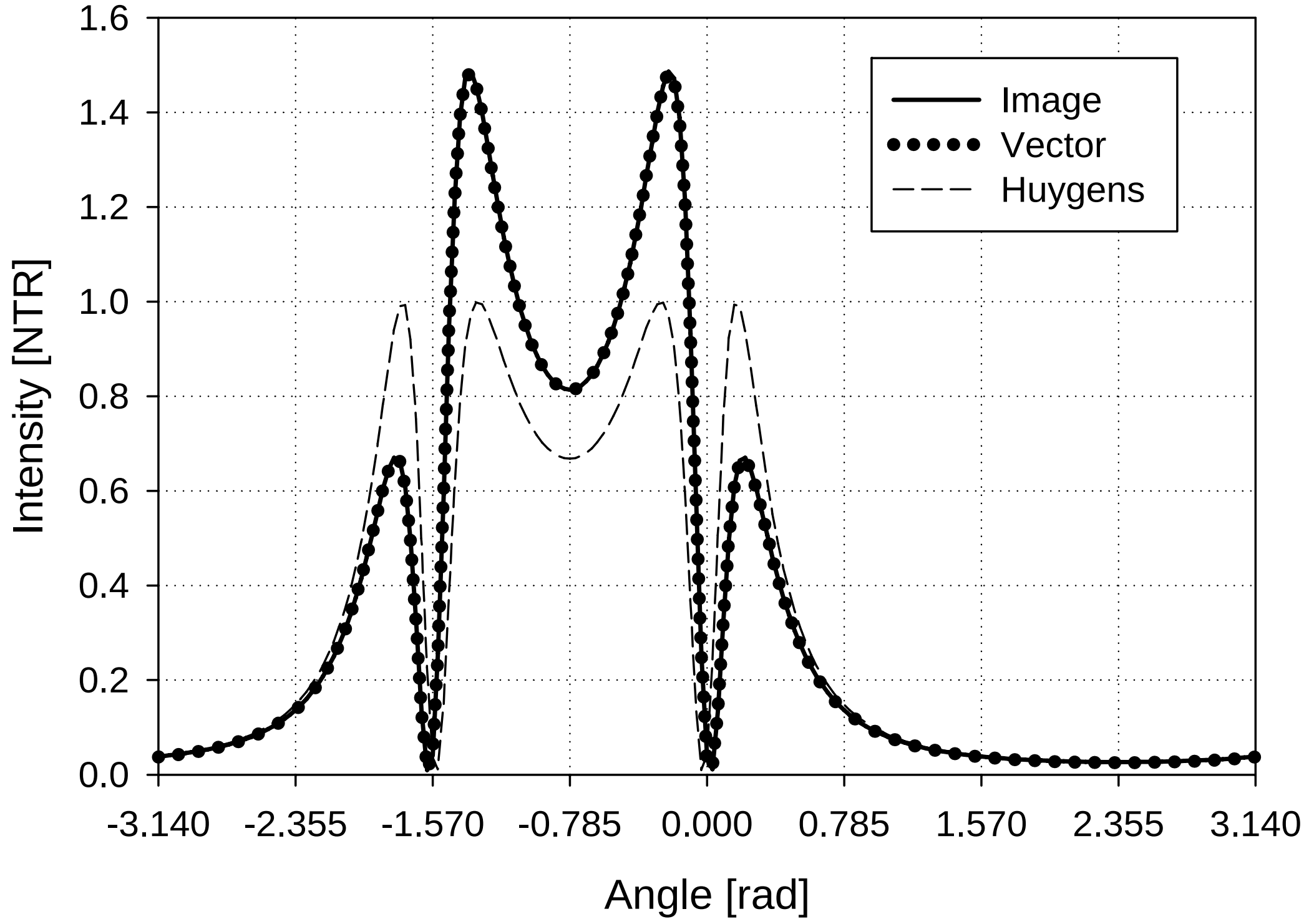

FIGURE 6

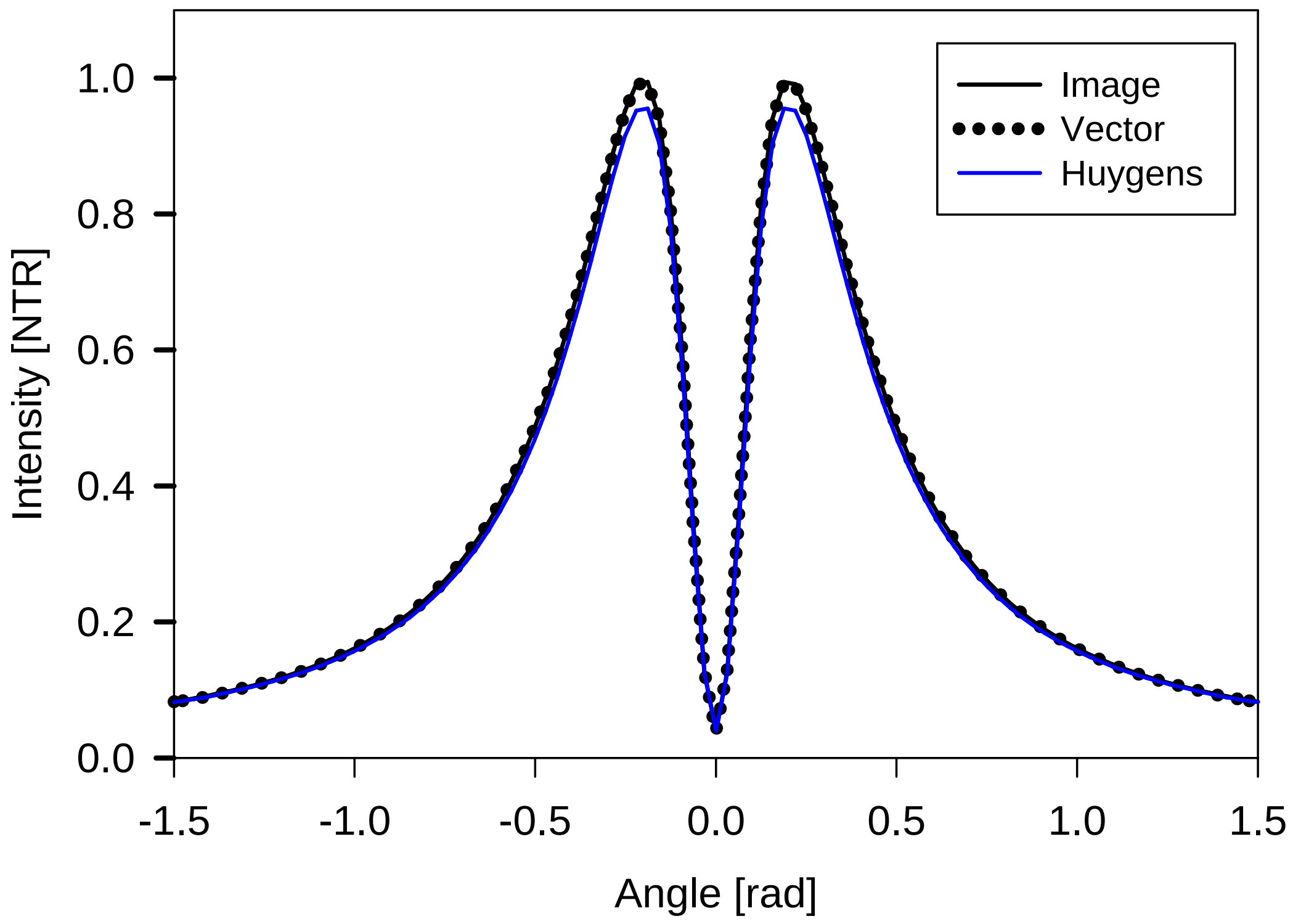

FIGURE 7

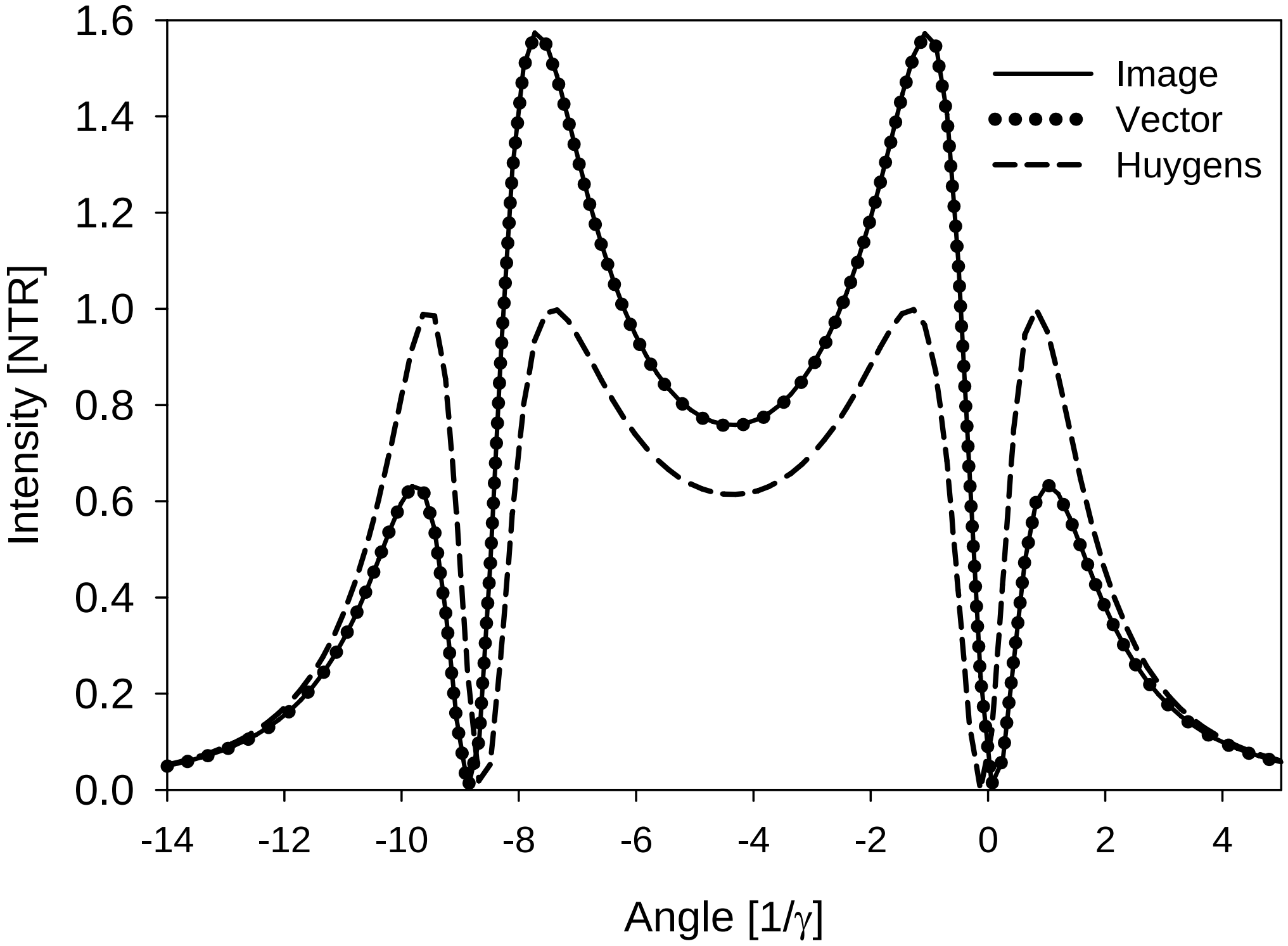

FIGURE 8

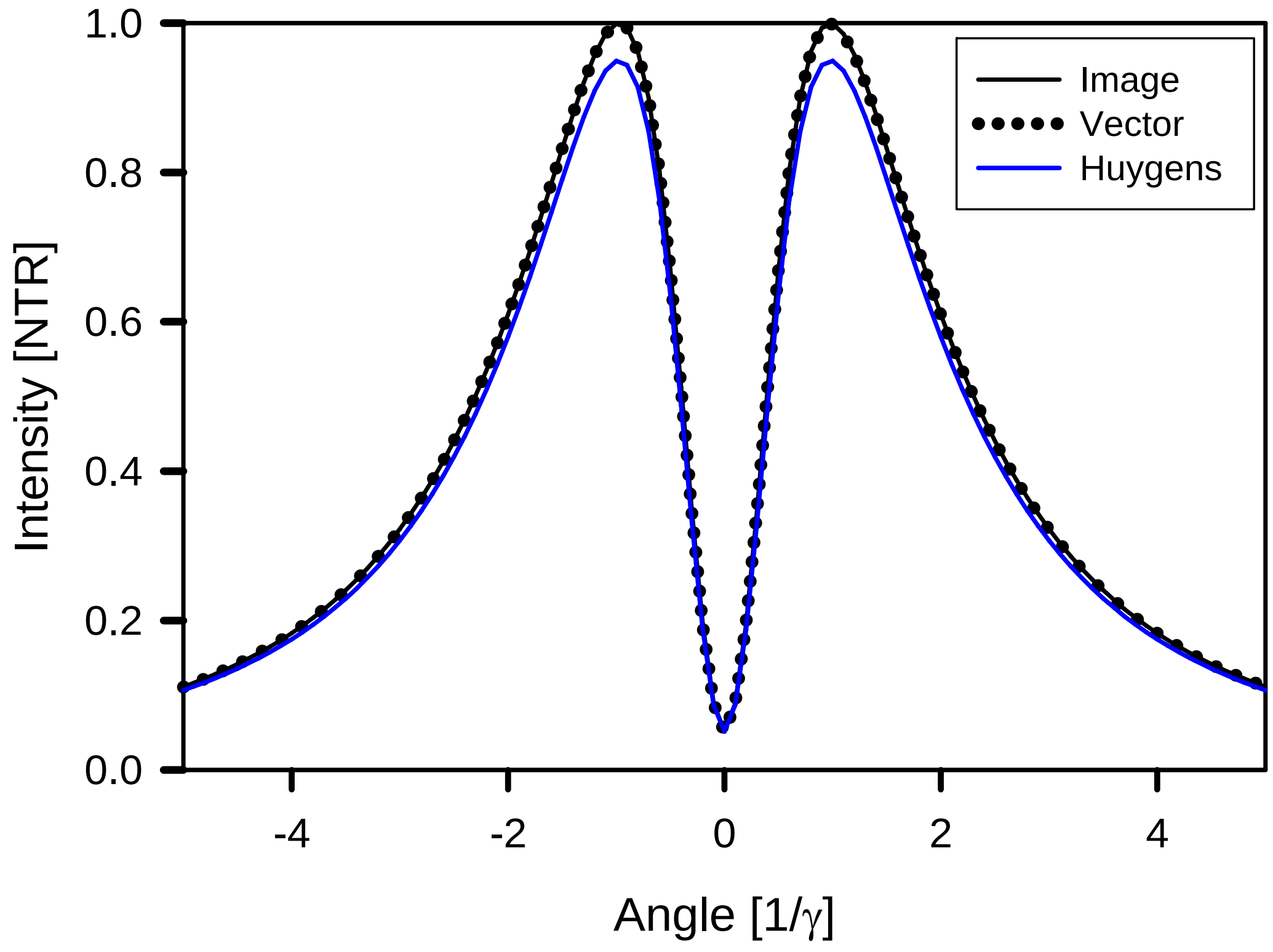

FIGURE 9

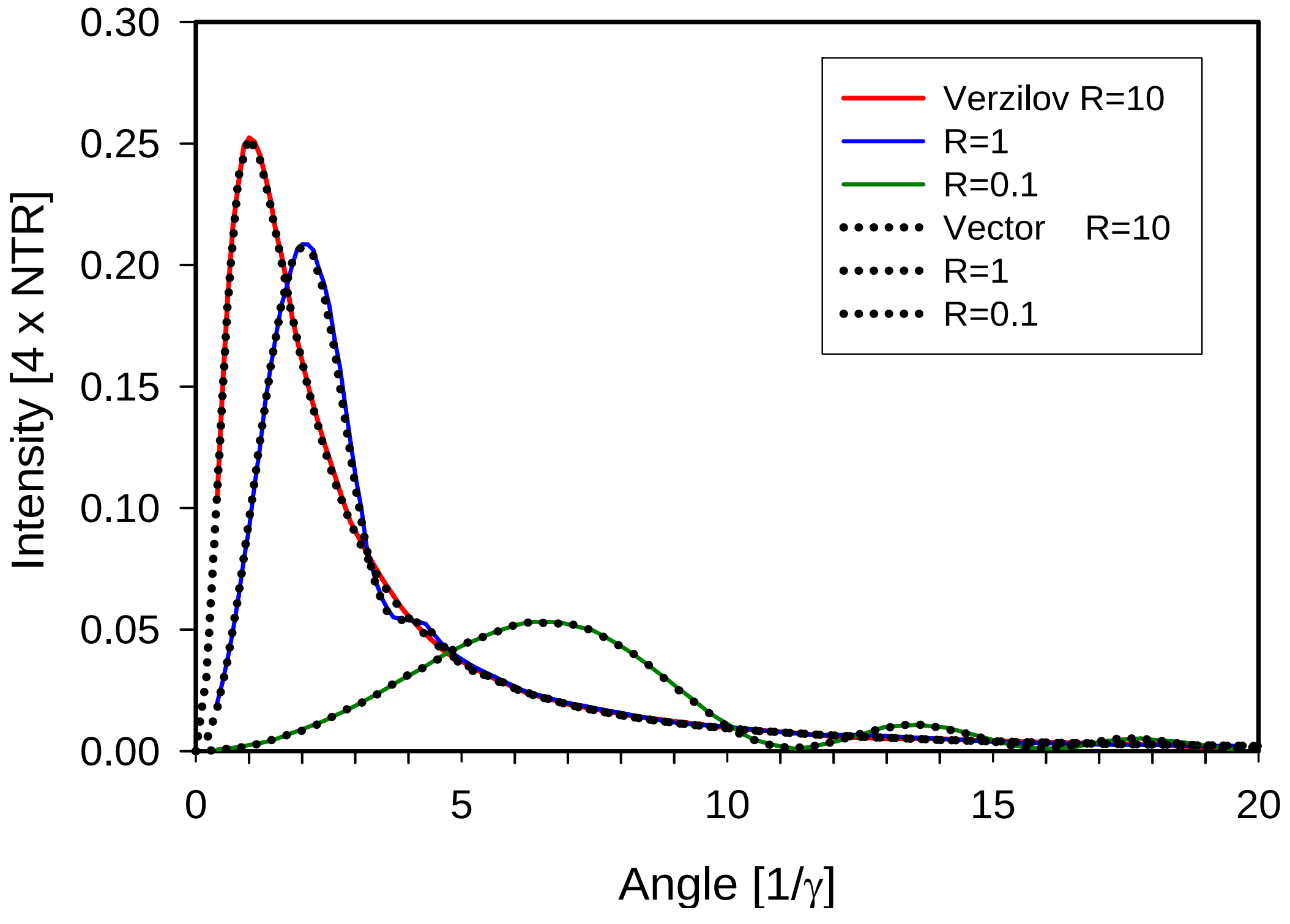

FIGURE 10

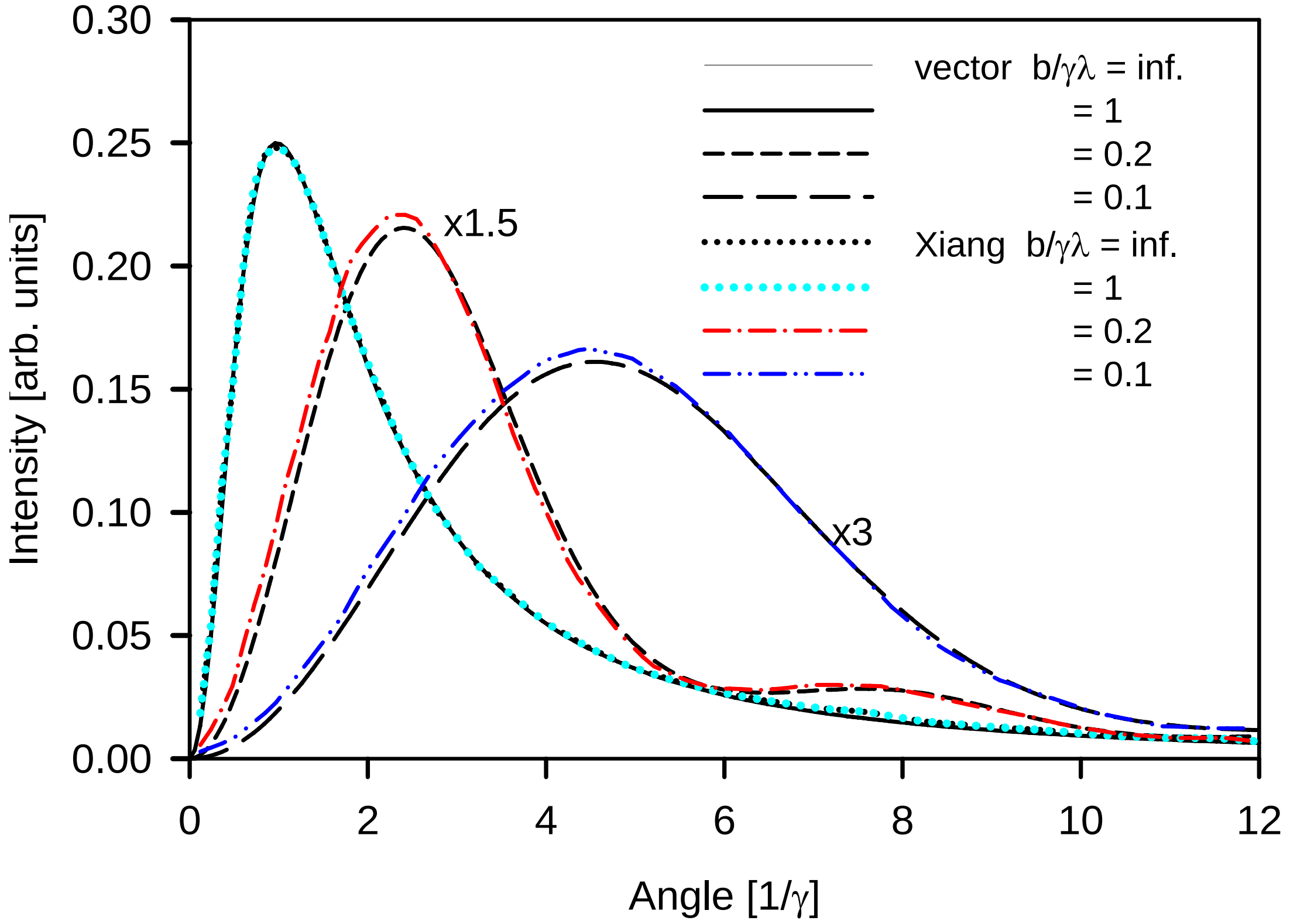

FIGURE 11

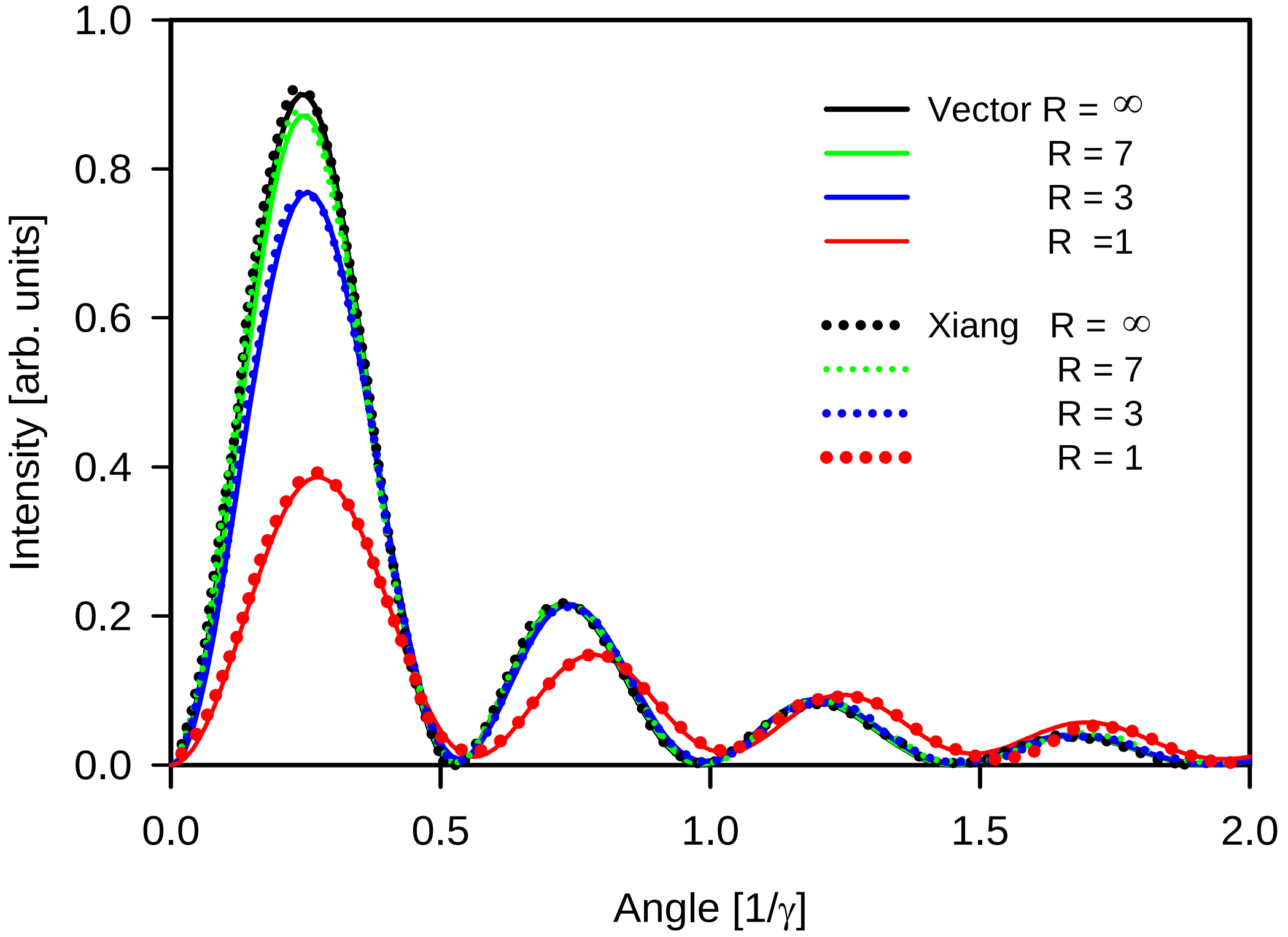

FIGURE 12

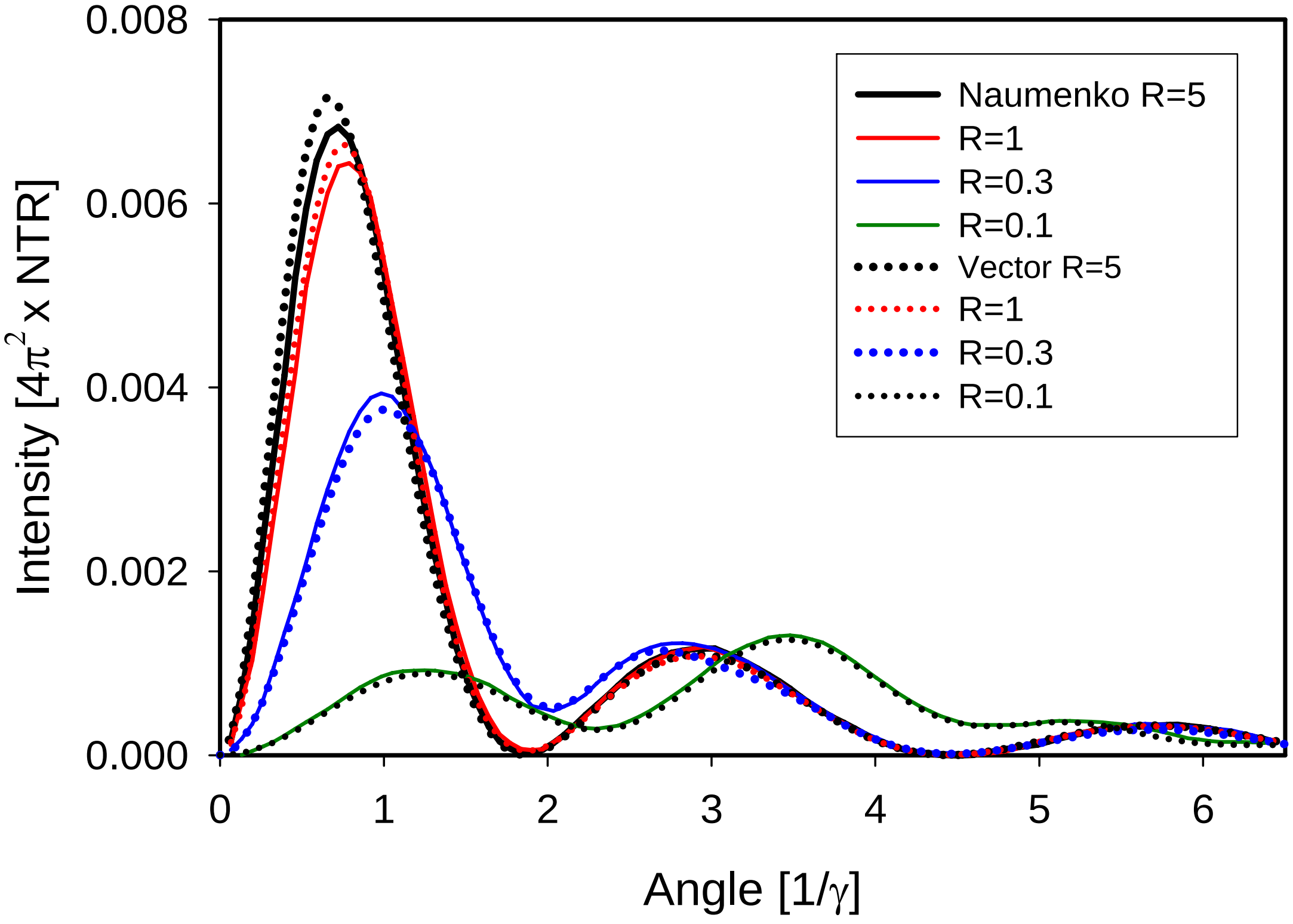

FIGURE 13

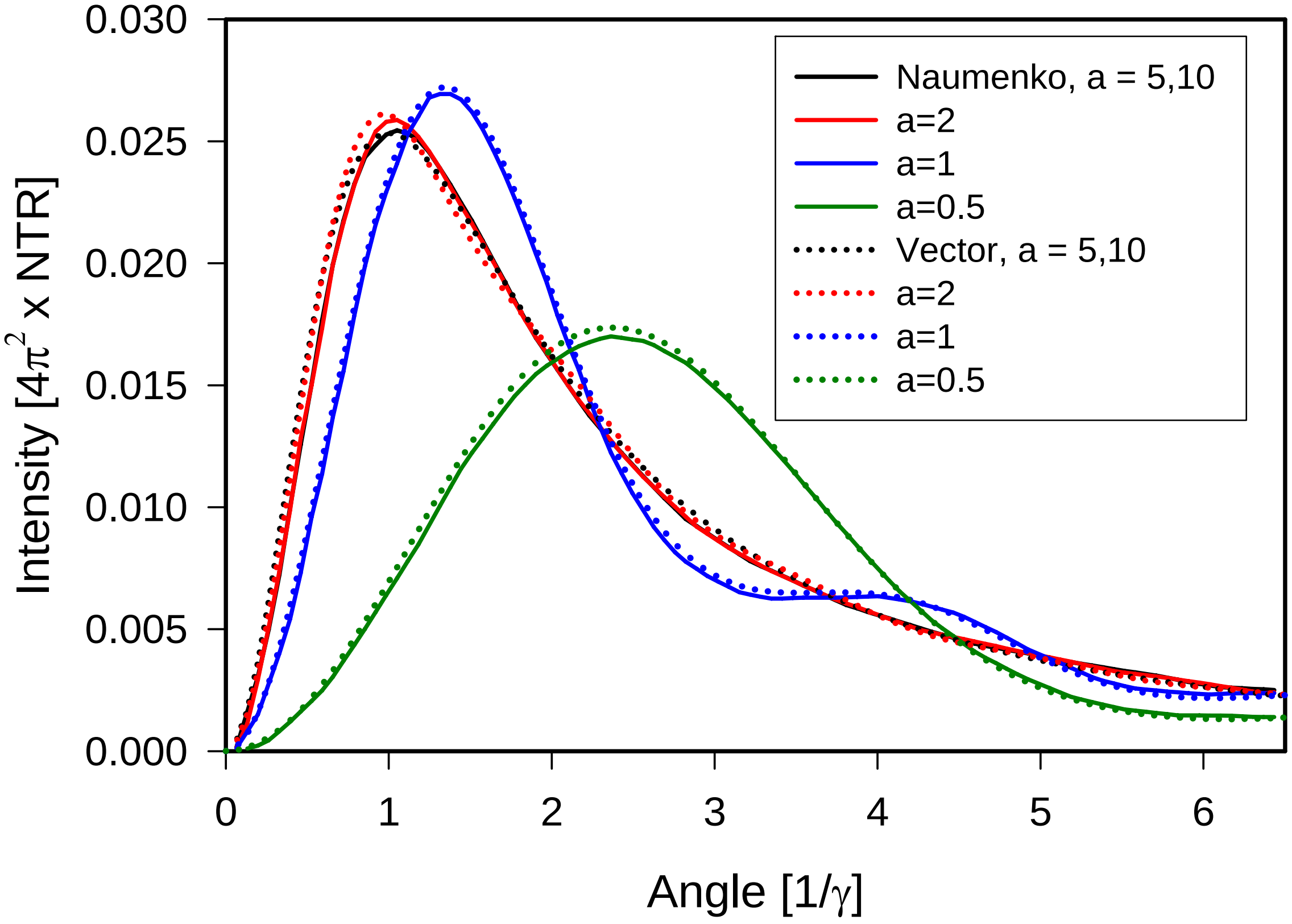

FIGURE 14

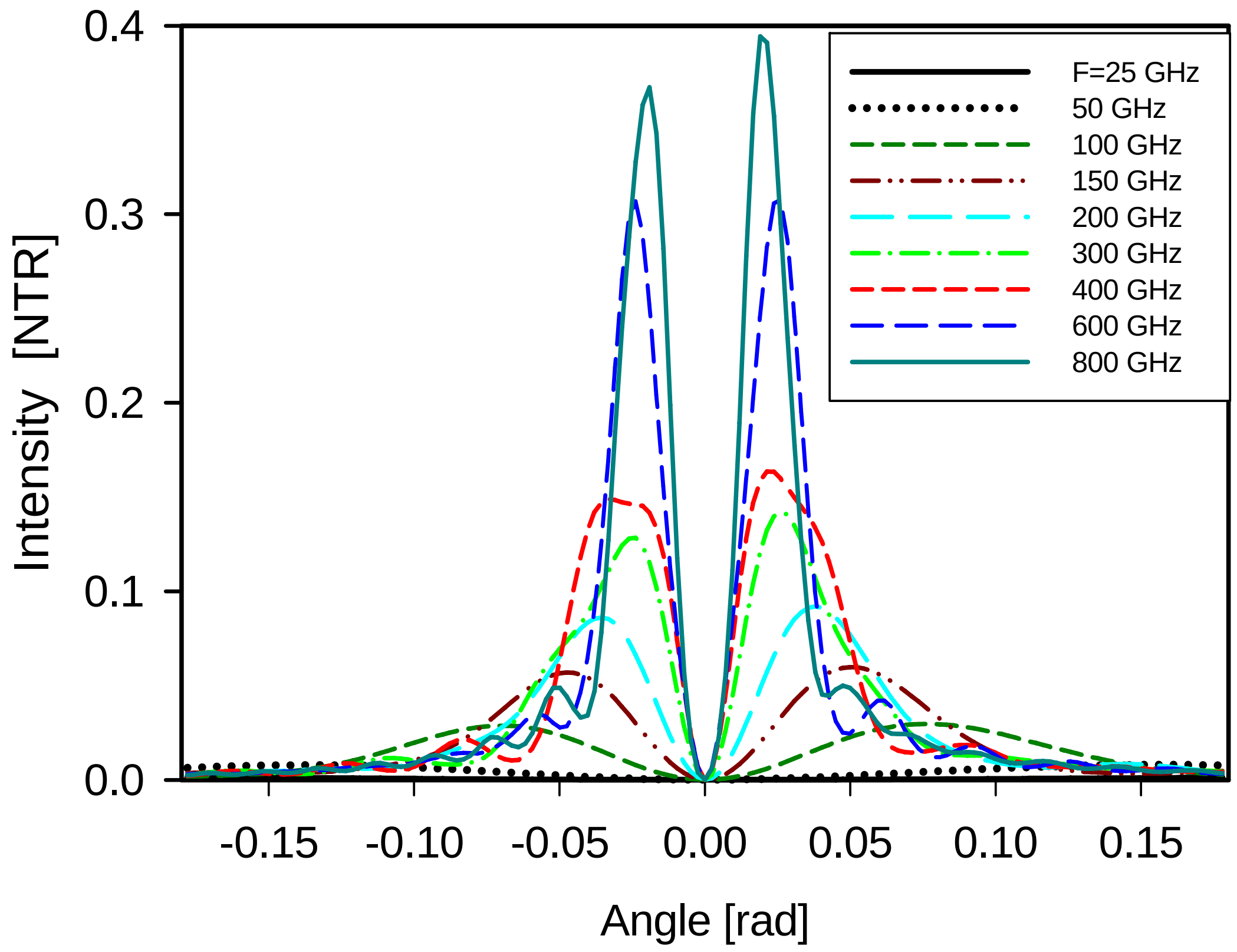

FIGURE 15

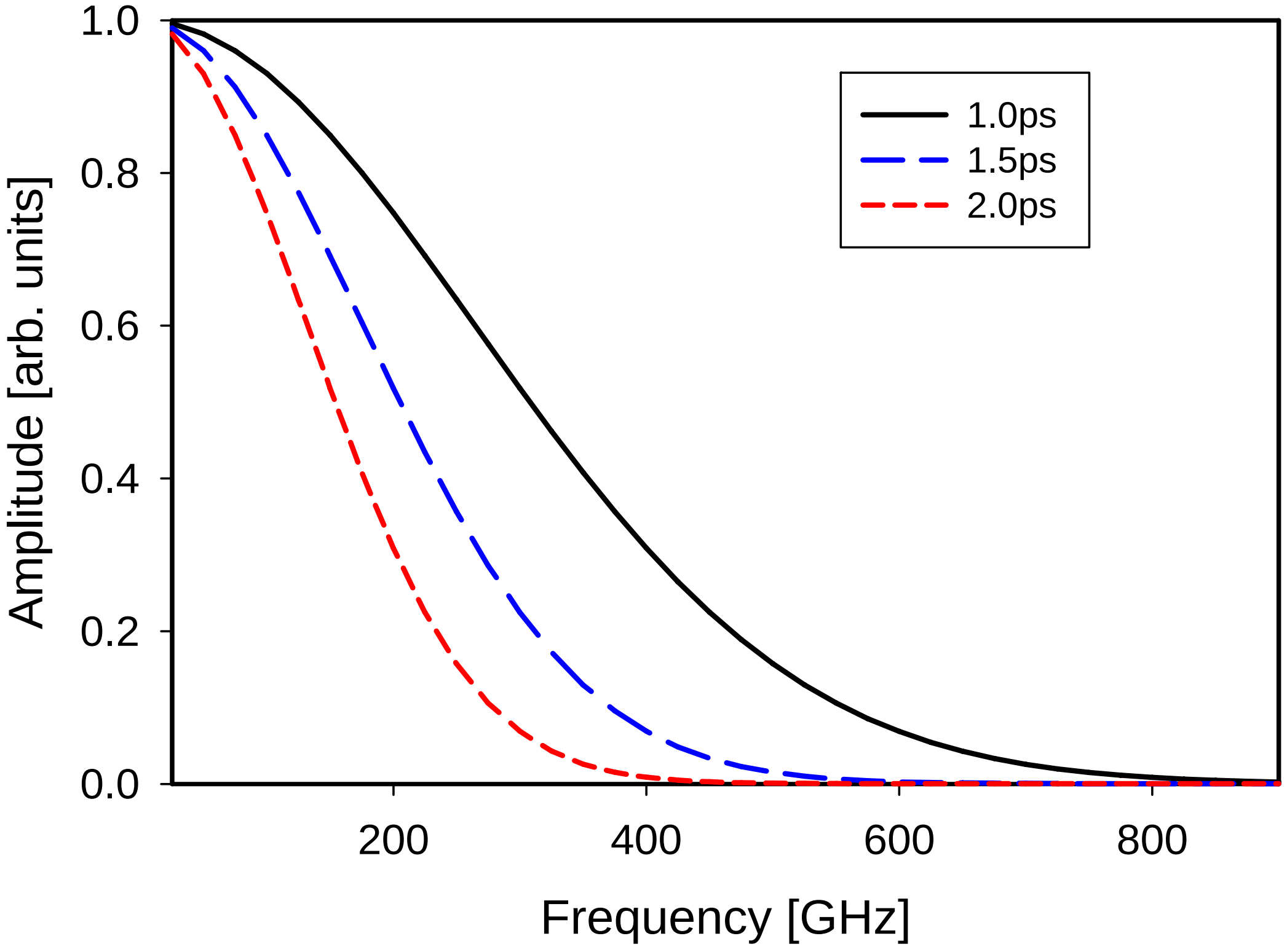

FIGURE 16

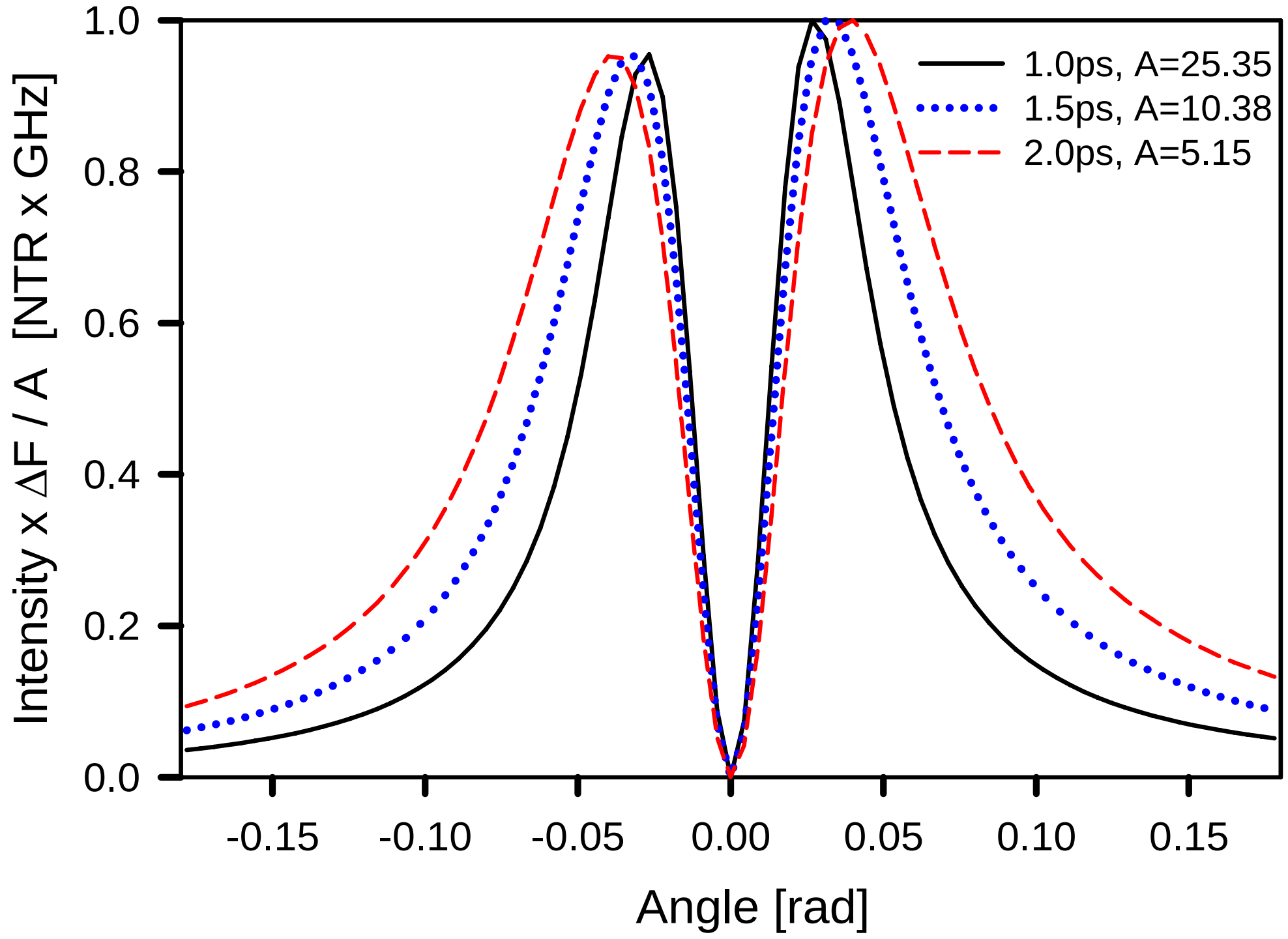
1.0
0.8
0.6
0.4
0.2
0.0
Intensity x ΔF / A [NTR x GHz]
-0.15
-0.10
-0.05
0.00
0.05
0.10
0.15
Angle [rad]
1.0ps, A=25.35
1.5ps, A=10.38
2.0ps, A=5.15